\documentclass[aps,prd,onecolumn,preprintnumbers,superscriptaddress,tightenlines,amsmath,amssymb,showpacs,nofootinbib]{revtex4-2}
\usepackage{textcomp}
\usepackage[english]{babel}
\usepackage{amsmath,amssymb,amsbsy,booktabs}
\usepackage{bm}
\usepackage{xcolor}
\usepackage{array}
\usepackage{amstext}
\usepackage{graphicx}
\usepackage{amsfonts}
\usepackage{bm}
\usepackage{dcolumn}
\usepackage{rotating}
\usepackage{epstopdf}
\usepackage{esint}
\usepackage{url}
\usepackage{pifont}
\usepackage{colortbl} 
\usepackage[colorlinks=true,
linkcolor=blue,
filecolor=blue,
anchorcolor=blue,
urlcolor=blue,
citecolor=blue
]{hyperref}

\usepackage[utf8]{inputenc}

\usepackage[T1]{fontenc} % Output font encoding for international characters

\usepackage{multirow}

\usepackage{array}
\usepackage{booktabs}
\usepackage{times}
\usepackage{diagbox}
\usepackage{slashed}
\usepackage{simpler-wick}

\allowdisplaybreaks[0]

\begin{document}

\title {\bf Radiative generation of chiral vector operators in $b\to s \nu\bar{\nu}$ transition}

\author{Xin-Shuai Yan}
\email{yanxinshuai@htu.edu.cn}
\affiliation{Institute of Particle and Nuclear Physics, Henan Normal University, Xinxiang, Henan 453007, China}

\author{Wen-Feng Liu}
\email{liuwenfeng2022@stu.htu.edu.cn}
\affiliation{Institute of Particle and Nuclear Physics, Henan Normal University, Xinxiang, Henan 453007, China}

\author{Qin Chang}
\email{changqin@htu.edu.cn}
\affiliation{Institute of Particle and Nuclear Physics, Henan Normal University, Xinxiang, Henan 453007, China}

\author{Ya-Dong Yang}
\email{yangyd@mail.ccnu.edu.cn}
\affiliation{Institute of Particle and Nuclear Physics, Henan Normal University, Xinxiang, Henan 453007, China}
\affiliation{Institute of Particle Physics and Key Laboratory of Quark and Lepton Physics~(MOE),\\
	Central China Normal University, Wuhan, Hubei 430079, China}

\begin{abstract}

The recent Belle II evidence for $B^+ \to K^+ \nu \bar{\nu}$, combined with a suppressed branching fraction ratio $R \equiv \mathcal{B}(B^0 \to K^{*0} \nu \bar{\nu}) / \mathcal{B}(B^+ \to K^+ \nu \bar{\nu})$, requires new physics that contributes to both left- and right-handed vector operators. 
We perform a systematic topological classification of one-loop completions that radiatively generate both operators without tree-level mediation, and construct two minimal benchmark scenarios: a scalar-rich model and a fermion-rich model. Evaluating these frameworks under specific benchmark mass schemes and four distinct flavor structures, we find a generic anti-correlation, where enhancing one decay channel typically suppresses the other. A notable exception is a decoupled flavor configuration within the scalar-rich model, which yields simultaneous constructive interference, reducing $R$ by $3.83\%$ while satisfying all complementary flavor bounds. Under a universal coupling assumption, the scalar-rich model yields exact cancellation in the charged mode, highlighting the non-trivial interplay between flavor structure and loop-generated Wilson coefficients.

\end{abstract}

\maketitle

\section{Introduction} 
\label{sec:intro}

Direct searches for new particles at the LHC have so far yielded null results, placing stringent constraints on physics beyond the Standard Model (SM). In contrast, indirect probes in the flavor sector exhibit persistent tensions, most notably in the flavor-changing neutral current (FCNC) transition $b \to s \nu \bar{\nu}$. The Belle II collaboration recently reported the first evidence for $B^+ \to K^+ \nu \bar{\nu}$ with a branching fraction of $\mathcal{B}_{\text{exp}} = (2.3 \pm 0.7) \times 10^{-5}$~\cite{Belle-II:2023esi}, exceeding the SM prediction of $(4.44 \pm 0.30) \times 10^{-6}$~\cite{Becirevic:2023aov} by $2.7\sigma$. This quark-level transition also governs the neutral mode $B^0 \to K^{*0} \nu \bar{\nu}$, where the current experimental upper limit of $1.8 \times 10^{-5}$~\cite{Belle:2017oht} lies roughly a factor of two above the SM expectation of $(9.47 \pm 1.40) \times 10^{-6}$~\cite{Becirevic:2023aov}.

These findings have stimulated a wide range of theoretical interpretations. One class of models postulates the existence of light, invisible states—such as sterile neutrinos~\cite{Browder:2021hbl,He:2021yoz,Felkl:2021uxi,Felkl:2023ayn,Ovchynnikov:2023von,Gartner:2024muk,Rosauro-Alcaraz:2024mvx,Buras:2024ewl,Becirevic:2024iyi,Datta:2023iln}, light dark matter (DM)~\cite{Bird:2004ts,Altmannshofer:2009ma,He:2022ljo,He:2023bnk,Hou:2024vyw,He:2024iju,Ho:2024cwk,Abdughani:2023dlr,Berezhnoy:2023rxx,Liu:2025lbw}, axion-like particles~\cite{Berezhiani:1989fs,Berezhiani:1990jj,Berezhiani:1990wn,Ferber:2022rsf,Altmannshofer:2023hkn,Altmannshofer:2024kxb,Dai:2024onu,Calibbi:2025rpx}, or generic long-lived particles~\cite{Filimonova:2019tuy,Fridell:2023ssf,Hu:2024mgf,Bolton:2024egx,Gabrielli:2024wys,McKeen:2023uzo,Davoudiasl:2024cee}—which mimic the missing energy signature of the neutrino pairs. Alternatively, the anomalies can be addressed within the framework of heavy new physics (NP), where virtual exchanges of heavy mediators modify the Wilson coefficients of the low-energy effective Lagrangian~\cite{Fuentes-Martin:2020hvc,Athron:2023hmz,Bause:2023mfe,Allwicher:2023xba,Chen:2023wpb,Chen:2024jlj,Hou:2024vyw,Marzocca:2024hua,Karmakar:2024gla,Hati:2024ppg,Allwicher:2024ncl,Tian:2024ubt,Alda:2024sup,Dev:2024tto,Becirevic:2024iyi,Buras:2024mnq,Bhattacharya:2024clv,Chen:2024cll,Kim:2024tsm,Zhang:2024hkn}. 

These tensions can be further quantified by the ratio $R \equiv \mathcal{B}(B^0 \to K^{*0} \nu \bar{\nu}) / \mathcal{B}(B^+ \to K^+ \nu \bar{\nu})$. The experimental upper limit, $R_{\mathrm{exp}} \le 1.125$, lies well below the SM prediction $ R_{\text{SM}} = 2.14 \pm 0.35$. In the framework of low-energy effective theory (LEFT), assuming the neutrinos remain purely left-handed, scalar and tensor contributions vanish due to chirality constraints. Consequently, the relevant interactions are governed by the vector operators:
\begin{align}\label{eq:LowOp}
	\mathcal{O}_V^{L} &= (\bar{s}\gamma_\mu P_L b)(\bar{\nu}_\ell\gamma^{\mu} P_L\nu_\ell)\,, \notag \\
	\mathcal{O}_V^{R} &= (\bar{s}\gamma_\mu P_R b)(\bar{\nu}_\ell \gamma^{\mu} P_L\nu_\ell)\,.	
\end{align}
Clearly, the tensions cannot be resolved by modifying only the left-handed operator $\mathcal{O}_V^L$, as such a modification would rescale both rates identically, leaving the ratio invariant. Thus, the presence of the right-handed operator $\mathcal{O}_V^R$ is a mandatory requirement for any explanation involving these vector currents~\cite{Bause:2023mfe,Chen:2024jlj}.

As highlighted in Ref.~\cite{Bause:2023mfe}, addressing the deviation in $R$ at tree level typically requires a non-minimal field content, such as the introduction of at least three distinct leptoquarks, while loop-level solutions have broadly been considered challenging. Nonetheless, the radiative generation of $\mathcal{O}_V^{L}$ and $\mathcal{O}_V^{R}$ presents a well-motivated model-building paradigm. Historically, loop-induced frameworks have provided economical pathways to unify disparate phenomenological puzzles---such as neutrino masses, dark matter, and other flavor anomalies---by linking them through a shared internal sector. Furthermore, loop-induced processes inherently provide a natural suppression mechanism. Should future, higher-statistics experimental updates from Belle II shift the central values and reduce the size of the anomaly, a radiatively generated contribution would perfectly accommodate a smaller, residual deviation without necessitating tiny tree-level couplings. Beyond these phenomenological considerations, systematically generating both left- and right-handed vector operators via loop processes represents a non-trivial structural feature of effective field theory. Yet, a comprehensive classification of such models remains systematically unexplored.

Motivated by these considerations, we employ a diagrammatic method to generate the complete set of one-loop topologies involving new scalars and fermions. We classify the minimal ultraviolet (UV) extensions capable of simultaneously generating both $\mathcal{O}_V^L$ and $\mathcal{O}_V^R$. To systematically navigate the multi-parameter space and explore how different coupling degeneracies govern the interference patterns in the decay rates, we define four representative flavor structures for the quark sector. By analyzing these specific arrangements, we quantitatively determine the maximum reach of these radiative contributions while ensuring compatibility with complementary flavor observables.

The remainder of this paper is organized as follows. In Sec.~\ref{sec:classification}, we present the systematic classification of one-loop topologies and identify the UV extensions capable of generating the required effective operators. Based on it, we select two minimal benchmark models for phenomenological study: one combines diagrams that independently generate $\mathcal{O}_V^R$ and $\mathcal{O}_V^L$; the other employs a single topology that produces both chiralities. Section~\ref{sec:benchmark-model} details these models. In Sec.~\ref{sec:Constraints}, we define the relevant observables and establish the constraints on the Wilson coefficients imposed by experimental data. The phenomenological analysis of these models is presented in Sec.~\ref{sec:Pheno}. Finally, we summarize our findings in Sec.~\ref{sec:Summary}, with technical details provided in the Appendices.

\section{Systematic Classification: from topologies to models}
\label{sec:classification}

In this work, our primary focus is on the radiative generation of the relevant effective operators. To systematically isolate and evaluate these loop-level effects, our phenomenological analysis proceeds under the working assumption that they provide the dominant new physics contribution.  This condition can be naturally enforced in various models through the imposition of a discrete or flavor symmetry in a manner analogous to established models of radiative neutrino mass generation (see e.g., Refs.~\cite{Yao:2018ekp,CentellesChulia:2019xky,Li:2022chc}). As the introduction of such symmetries is a standard technique in model-building, we proceed with the assumption that a suitable symmetry can always be implemented to forbid the tree-level contributions. Consequently, we will not specify the exact symmetry structure for each case but will instead focus on the general classification of all valid one-loop model structures.

Our classification scheme is built upon a hierarchy of concepts~\cite{CentellesChulia:2019xky}, which we define as follows:
\begin{itemize}
	\item \textbf{Topology}: We define a \textit{topology} as the underlying graphical structure of a Feynman diagram, abstracting from the physical properties of the particles involved. It is defined solely by its vertices and their connectivity, without consideration for the Lorentz nature or quantum numbers of the fields.
	
	\item \textbf{Renormalizable Topology}: A topology is deemed \textit{renormalizable} if it can be constructed exclusively from vertices corresponding to interactions of mass dimension less than or equal to four. This ensures that the corresponding operator can be embedded within a renormalizable quantum field theory.
	
	\item \textbf{Diagram}: We refer to a topology as a \textit{diagram} once a specific particle type (i.e., scalar, vector, or fermion) has been assigned to each line. 
	
	\item \textbf{Model-Diagram}: A \textit{model-diagram} is a fully specified diagram where all internal fields are assigned a complete set of quantum numbers under relevant symmetries (e.g., SM gauge group representations).
\end{itemize}

\begin{figure}[t]
	\centering
	\includegraphics[width=0.7\textwidth]{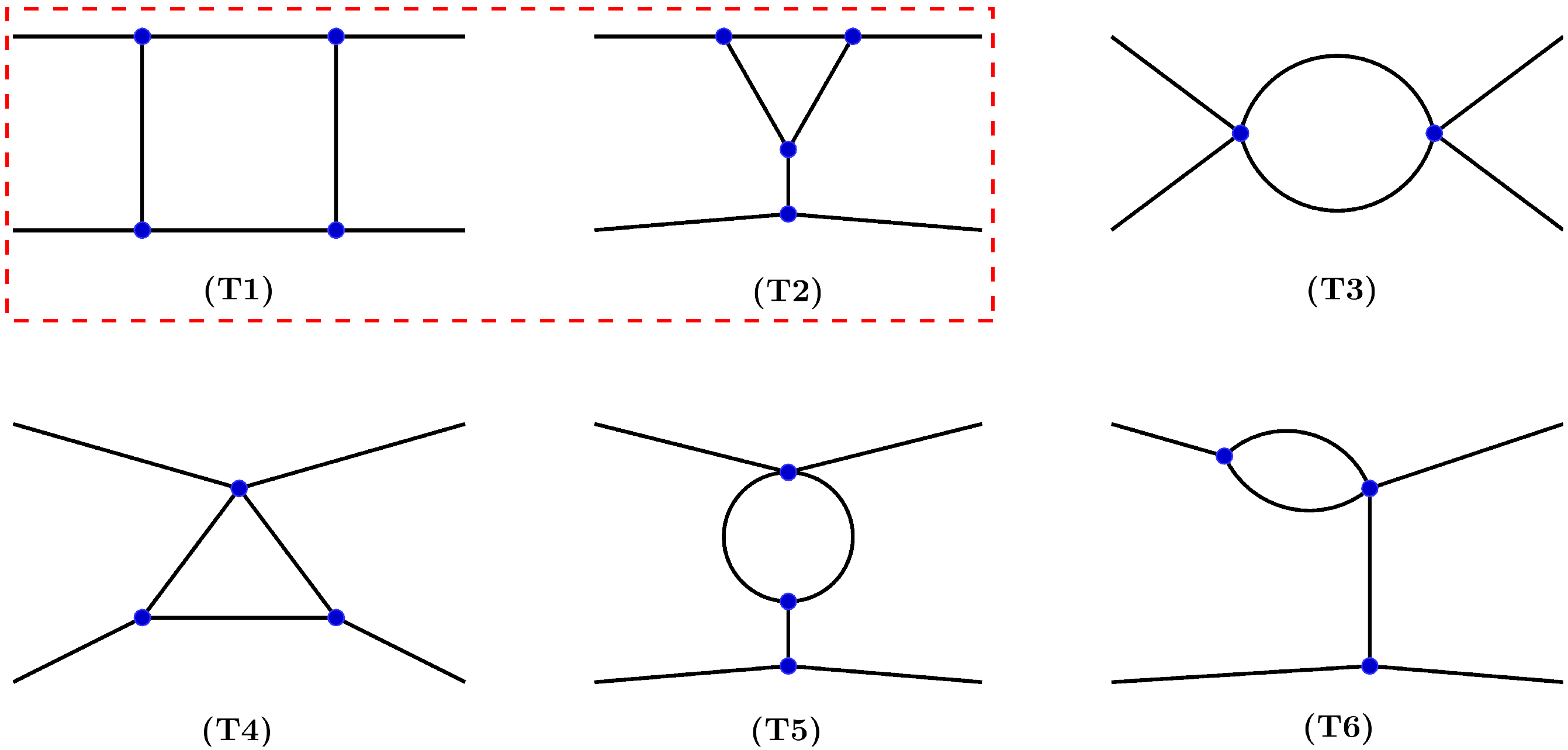}
	\caption{One-loop topologies with 3- and 4-point vertices and four external legs. Topologies T1 and T2, highlighted in the red-dashed box, are renormalizable.}
	\label{fig:topology} 
\end{figure}

We begin our classification by systematically generating all one-loop topologies for a four-point interaction using only three- and four-point vertices. From this initial set, we discard all graphs corresponding to tadpoles or external-line self-energy corrections, as they lead to divergent loop integrals. 
This leaves six topologies, which we depict in Fig.~\ref{fig:topology}. 

We categorize these six topologies according to the definitions established previously. The first class, comprising T1 and T2, is composed exclusively of three-point vertices and is therefore classified as \textit{renormalizable}. In contrast, the remaining topologies—T3, T4, T5, and T6—each contain at least one four-point vertex. Given that all external lines in our operators are fermionic, any such four-point vertex necessarily corresponds to an operator with a mass dimension greater than four. Consequently, these latter four topologies are classified as \textit{non-renormalizable} and will not be considered further in this work.

\begin{figure}[t]
	\centering
	\includegraphics[width=0.7\textwidth]{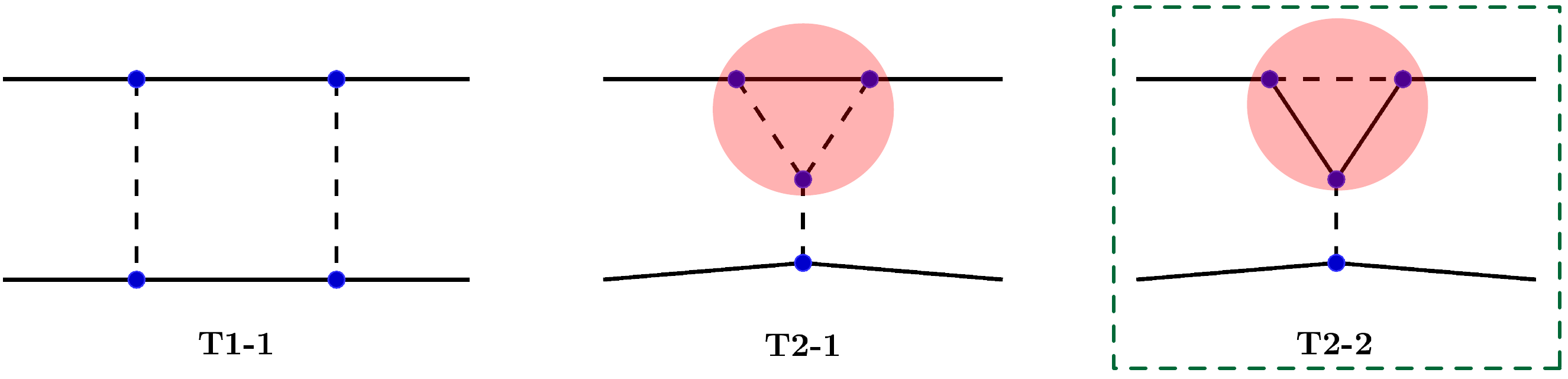}
	\caption{Possible one-loop diagrams for the topologies T1 and T2, where the dashed lines denote scalars and the solid lines denote fermions. 
		All the diagrams for T2 contain a reducible 3-point loop vertex, highlighted in red. The diagram T2-2, enclosed in the green-dashed box, 
		possesses an ultraviolet divergent loop integral. 
		 }
	\label{fig:diagrams} 
\end{figure}

We now proceed to construct the diagrams for the renormalizable topologies T1 and T2. By assigning fermion and scalar identities to the internal lines while keeping the four external lines as fermions, we obtain the three distinct diagrams shown in Fig.~\ref{fig:diagrams}. In our convention, solid lines denote fermions and dashed lines denote scalars. These diagrams are distinguished by their structure, which we categorize as either reducible or irreducible. Here we deem a one-loop diagram \textit{reducible} if it functions as a radiative correction to a renormalizable vertex.  Accordingly, diagram T1-1 is an irreducible box diagram, while T2-1 and T2-2 are reducible, each containing a one-loop vertex correction as highlighted in red.

This structural distinction has a critical physical implication. The presence of a reducible vertex correction, as seen in T2-1 and T2-2, generically implies the existence of a corresponding tree-level diagram that contributes to the same four-point operator. To ensure the one-loop contribution is dominant without fine-tuning or the imposition of special symmetries to forbid the tree-level term, such diagrams must be set aside. Furthermore, the loop integral in diagram T2-2 (enclosed in the green-dashed box) is UV divergent. For these reasons, we will focus our subsequent analysis on the irreducible T1-1 diagram, which represents a \textit{genuine} one-loop realization~\cite{Yao:2018ekp,CentellesChulia:2019xky}. We comment, however, that diagram T2-1 remains a viable candidate in models where a symmetry is imposed to forbid its tree-level contribution (e.g., under a $Z_2$ symmetry where the SM fields and the NP mediator connecting the two external lines are $Z_2$-even, while the three NP fields running in the loop are $Z_2$-odd).

\begin{figure}[t]
	\centering
	\includegraphics[width=0.6\textwidth]{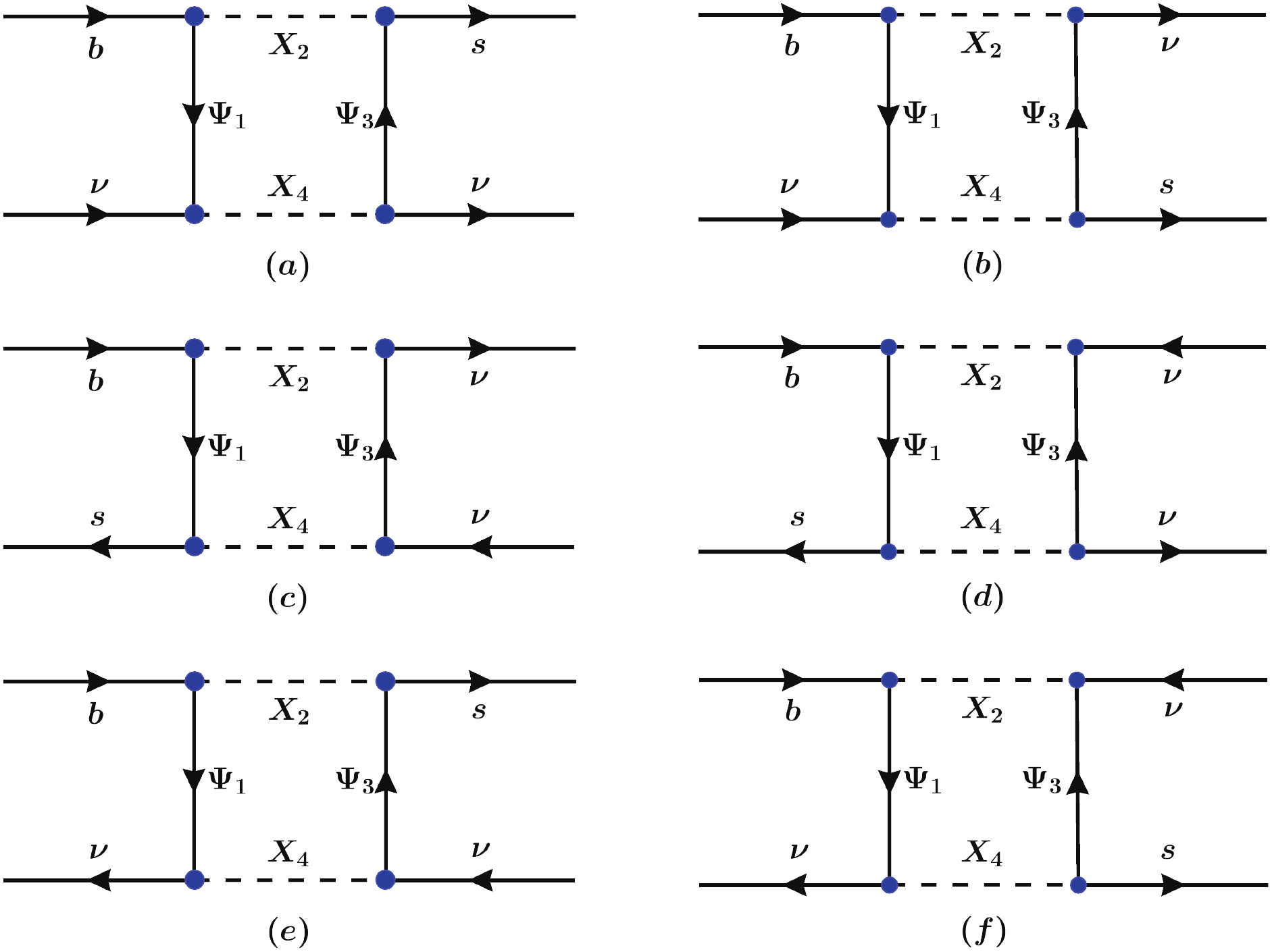}
	\caption{The six distinct diagrams that can be constructed from the T1-1 diagram by permuting the external fields ($b, s, \nu, \nu$). The internal fields are fixed as fermions ($\Psi_1, \Psi_3$) and scalars ($X_2, X_4$).}
	\label{fig:box} 
\end{figure} 

\begin{table}[!h] 
	\renewcommand\arraystretch{1.3}  
	\centering 
	\tabcolsep=0.38cm
    \caption{SM gauge quantum number assignments for the internal fields of diagram ($c$) that generate the operator $\mathcal{O}^L_V$. The SU(3)$_c$ and SU(2)$_L$ representations are presented as distinct cases, while the U(1)$_Y$ hypercharges are fixed relative to a single free parameter, $\alpha = Y_{\Psi_1}$.}
	\begin{tabular}{c|cccc}
		\hline \hline
		SU(3)$_c$ & $\Psi_1$    & $X_2$  & $\Psi_3$ & $X_4$  \\
		\hline
		A & 1 & $\bar{3}$ & $\bar{3}$ & $\bar{3}$ \\
		B & 3 & 1 & 1 & 1 \\
		\hline \hline
		SU(2)$_L$ & $\Psi_1$    & $X_2$  & $\Psi_3$ & $X_4$  \\
		\hline
		I & 1 & 2 & 3 & 2 \\
		II & 2 &1 & 2 & 1 \\
		III & 2 & 1 & 2 &3 \\
		IV & 3 & 2& 1 & 2\\
		V & 3 & 2 & 3 & 2\\
		VI & 2 &3 & 2 & 1 \\
		VII & 2 & 3 & 2 &3 \\
		\hline \hline
		U(1)$_Y$ & $\Psi_1$    & $X_2$  & $\Psi_3$ & $X_4$  \\
		\hline
		 i & $\alpha$ & $\alpha-1/6$ & $\alpha-2/3$ & $\alpha-1/6$ \\
		\hline \hline
	\end{tabular}
	\label{tab:T1-c-L}   
\end{table}

With the T1-1 diagram established as our template, the next step is to construct its concrete physical realizations. We assign a specific particle content to the internal lines: two vector-like fermions ($\Psi_1, \Psi_3$) and two scalars ($X_2, X_4$). We then attach the external fields—a $b$ quark, an $s$ quark, and two neutrinos ($\nu$)—in all non-equivalent configurations, leaving their chirality unspecified at this stage. This procedure yields the six unique diagrams depicted in Fig.~\ref{fig:box}. For notational simplicity, we drop the ``T1-1'' prefix and will henceforth refer to these diagrams by their labels, ($a$) through ($f$).

In principle, the T1-1 topology can accommodate SM particles (such as SM fermions and electroweak 
gauge bosons) as well as exotic vector mediators. Here, we restrict all internal propagators 
to beyond-the-SM states, leaving open the possibility that these mediators belong to a dark or hidden 
sector. We focus exclusively on new scalars and fermions, omitting new vector bosons since their 
inclusion would typically require an extended gauge symmetry and an associated symmetry-breaking sector, 
introducing additional model-dependent constraints beyond our minimal setup. Finally, we take the new 
fermions to be vector-like under $\mathrm{SU}(3)_c \times \mathrm{SU}(2)_L \times \mathrm{U}(1)_Y$, 
which automatically preserves the gauge anomaly cancellation of the SM.

We now promote the diagrams to \textit{model-diagrams} by assigning SM quantum numbers to the internal fields, requiring SM gauge symmetry conservation at each vertex. The quantum numbers of the external fields provide the necessary boundary conditions, fixed by the choice of target operator ($\mathcal{O}_V^{R}$ or $\mathcal{O}_V^{L}$). For operators involving SU(2)$_L$ charges, we note that the external left-handed fermions shall be treated as components of their respective doublets: the $b$ and $s$ quarks as part of the quark doublet $Q$, and the neutrino as part of the lepton doublet $L$. The nature of the constraints on the internal fields then depends on the gauge group. For U(1)$_Y$, the additive nature of hypercharge means that a choice for one internal field's hypercharge fixes all others. For SU(3)$_c$ and SU(2)$_L$, however, the tensor product of representations decomposes into multiple irreducible representations, allowing for more freedom. In this work, to ensure a tractable classification, we restrict our analysis to color singlet and triplet for SU(3)$_c$, and SU(2)$_L$ representations up to triplet.

As a representative example, we detail in Table~\ref{tab:T1-c-L} the complete set of allowed SM quantum number assignments for the internal fields of diagram ($c$) that generate the operator $\mathcal{O}^L_V$. The analogous results for the remaining five diagrams are compiled in Appendix~\ref{app:models}. The table is organized into three sections, listing the viable assignments for the SU(3)$_c$ and SU(2)$_L$ representations, and the corresponding U(1)$_Y$ hypercharges. 
Each physical model is constructed by selecting one complete set of SU(3)$_c$ representations (e.g., case A) and one set of SU(2)$_L$ representations (e.g., case I). The hypercharges are not independent; as shown in the final section, they are all uniquely determined by the choice of a single free parameter, $\alpha$, which we define as the hypercharge of the fermion $\Psi_1$ ($\alpha \equiv Y_{\Psi_1}$).

A crucial aspect of this analysis is the matching procedure between the high-scale UV model and the low-energy effective operators. When integrating out the heavy internal fields, the UV model does not directly generate the LEFT operators $\mathcal{O}^{L,R}_V$. Instead, at the new physics scale $\Lambda$, it must first be matched onto the appropriate set of dimension-six operators in the Standard Model Effective Field Theory (SMEFT). For the processes we consider, the relevant SMEFT Lagrangian is
\begin{align}
	\mathcal{L}_{\text{SMEFT}} \supset  [C_{\ell d}]_{prst}Q_{\ell d}+ [C_{\ell q}^{(1)}]_{prst} Q^{(1)}_{\ell q} 
	+ [C_{\ell q}^{(3)}]_{prst} Q^{(3)}_{\ell q} \,,
\end{align}
where the operators are defined as
\begin{align}
	Q_{\ell d}&=(\bar{L}_p\gamma^\mu L_r)(\bar{d}_{Rs}\gamma_\mu d_{Rt})\,, \nonumber \\
	Q^{(1)}_{\ell q}&=(\bar{L}_p\gamma^\mu L_r)(\bar{Q}_s\gamma_\mu Q_t) \,, \nonumber \\
	Q^{(3)}_{\ell q}&=(\bar{L}_p\gamma^\mu \tau^I L_r)(\bar{Q}_s\gamma_\mu \tau^I Q_t)\,,
\end{align}
with $\tau^I$ being the Pauli matrices and $p, r, s, t$ the generation indices. 
In our calculation, we adopt two primary conventions. First, the new physics scale $\Lambda$ is absorbed into the definition of the Wilson coefficients. Second, we work in the ``down basis,'' where the down-type quark and charged-lepton Yukawa matrices are diagonal. Within this framework, the SMEFT operators must then be evolved via renormalization group (RG) running from the high scale down to the electroweak scale, where they can be matched onto the LEFT operator basis, which includes $\mathcal{O}^L_V$ and $\mathcal{O}^R_V$~\cite{Jenkins:2017jig}.

To illustrate this procedure, we perform an explicit calculation for a representative case: model A-V, constructed from diagram ($c$) using the SU(3)$_c$ and SU(2)$_L$ assignments from cases A and V in Table~\ref{tab:T1-c-L}, respectively. For simplicity, we assume that the scalars $X_2$ and $X_4$ are the same particle, as they share the same SM quantum numbers. The interaction Lagrangian for this model is given by
\begin{align}\label{eq:L-c}
	\mathcal{L} &\supset g^t_{1}\bar{\Psi}_{1}Q_{t} X_{2}+g^{p}_{2}\bar{L}_{p}\Psi_{3} X^{\dagger}_{2}
	+g^{r*}_{2}\bar{\Psi}_{3} L_{r}X_{2}+g^{s*}_{1}\bar{Q}_s \Psi_{1} X^{\dagger}_{2} \nonumber \\
	&= g^t_{1}\bar{\Psi}^{A}_{1}Q_{t\alpha} X_{2\alpha'}\xi^A_{\alpha\alpha'}+g^{p}_{2}\bar{L}_{p\beta}\Psi^B_{3} X^{*}_{2\beta'}(\xi^{B\dagger})_{\beta\beta'}
	+g^{r*}_{2}\bar{\Psi}^{A'}_{3} L_{r\rho}X_{2\rho'}\xi^{A'}_{\rho\rho'}
	+g^{s*}_{1}\bar{Q}_{s\sigma} \Psi^{B'}_{1} X^{*}_{2\sigma'}(\xi^{B'\dagger})_{\sigma\sigma'}\,,
\end{align} 
where in the second line we have made the SU(2)$_L$ structure explicit. The Greek letters ($\alpha, \beta, \dots$) are used to denote SU(2)$_L$ indices in the fundamental representation, taking values 1 and 2, while color indices are suppressed due to the trivial SU(3)$_c$ structure of this model. The letters $g_{1,2}$ denote the coupling constants, and the SU(2)$_L$ tensor $\xi$ is defined as $\xi^A=((1+\tau^3)/2, \tau^1/\sqrt{2},(1-\tau^3)/2)$ with $A\in 1,2,3$.

Integrating out the heavy fields $\Psi_i$ and $X_2$ at one loop yields the following effective four-fermion operator:
\begin{align}\label{eq:intermediate_result}
	\mathcal{L}_{\text{eff}} \supset
	(g^{t}_{1}g^{p}_{2}g^{r*}_{2}g^{s*}_{1})J_4(m_{X_2},m_{X_2},m_{\Psi_1},m_{\Psi_3})
	\Big(\bar{L}_{p\beta}\gamma^{\mu}L_{r\rho}\Big)
	\Big(\bar{Q}_{s\sigma}\gamma_{\mu}Q_{t\alpha}\Big)
	\xi^A_{\alpha\alpha'}(\xi^{B\dagger})_{\beta\alpha'}
	(\xi^{A\dagger})_{\sigma\rho'}\xi^{B}_{\rho\rho'}\,,
\end{align}	
where $J_4$ is the loop function defined in Appendix~\ref{app:loopfunc}. 
The final step is to evaluate the product of the SU(2)$_L$ tensors in Eq.~\eqref{eq:intermediate_result}. This algebraic simplification directly maps the operator onto a specific linear combination of the SMEFT operators $Q^{(1)}_{\ell q}$ and $Q^{(3)}_{\ell q}$. The final effective Lagrangian at the scale $\Lambda$ is therefore
\begin{align}
	\mathcal{L}_{\text{eff}}=\frac{1}{8} (g^{t}_{1}g^{p}_{2}g^{r*}_{2}g^{s*}_{1}) J_4(m_{X_2},m_{X_2},m_{\Psi_1},m_{\Psi_3})
	\left(9[Q^{(1)}_{\ell q}]_{prst}+[Q^{(3)}_{\ell q}]_{prst}\right)\,.
\end{align}	
The Wilson coefficients for the SMEFT operators are thus determined by the couplings and masses of the UV model. For the convenience of future reference, we list the coefficients of $Q^{(1)}_{\ell q}$ and $Q^{(3)}_{\ell q}$ for all valid models in Table~\ref{tab:T1-c-L-Mod}.

Two key features characterize this classification. First, certain gauge-invariant assignments are phenomenologically trivial. For instance, the $SU(2)_L$ assignment $(\Psi_1, X_2, \Psi_3, X_4) \to (1, 2, 1, 2)$ generates an $SU(2)_L$ structure of the form
\begin{align}
	\Big(\bar{L}_{p\alpha}\gamma^{\mu}L_{r\beta}\Big)
	\Big(\bar{Q}_{s\beta}\gamma_{\mu}Q_{t\alpha}\Big)
	=\frac{1}{2}\left([Q^{(1)}_{\ell q}]_{prst}+[Q^{(3)}_{\ell q}]_{prst}\right)\,.
\end{align}
Following electroweak symmetry breaking, the tree-level matching of this specific operator combination onto the target $\mathcal{O}^L_V$ vanishes. Consequently, we exclude solutions yielding this null result from our final analysis, as presented in Table~\ref{tab:T1-c-L-Mod} and Appendix~\ref{app:models}.

Second, a correspondence emerges between diagram topology and operator chirality in models with complex mediators (i.e., Dirac fermions and complex scalars). In this scenario, diagrams ($c$) and ($d$) are versatile, generating both $\mathcal{O}^L_V$ and $\mathcal{O}^R_V$, whereas diagrams ($a$) and ($b$) are restricted to $\mathcal{O}^R_V$, and ($e$) and ($f$) are limited to $\mathcal{O}^L_V$. This chiral distinction is relaxed in the presence of real fields, such as Majorana fermions or real scalars; these introduce new interaction structures that allow previously restricted diagrams to generate both chiralities. These effects are explicitly demonstrated via our benchmark models in Sec.~\ref{sec:benchmark-model}.

\begin{table}[t] 
	\renewcommand\arraystretch{1.6}  
	\centering 
	\tabcolsep=0.6cm % Adjusted spacing for clarity
   \caption{Relative strengths of the SMEFT operators $Q^{(1)}_{\ell q}$ and $Q^{(3)}_{\ell q}$ generated by diagram ($c$) for the different SU(2)$_L$ models. An overall normalization factor, which includes the loop function and couplings, is not presented.}
	\begin{tabular}{c|ccccccc} 
		\hline\hline
		\diagbox[width=7.5em]{Operator}{ Model} & I & II & III & IV & V & VI & VII \\
		\hline
		$Q^{(1)}_{\ell q}$ &  3/4 & 1 &  0  &  3/4 & 9/8 &  0  & 3/4 \\
		$Q^{(3)}_{\ell q}$ & -1/4 & 0 & 1/2 & -1/4 & 1/8 & 1/2 & 1/2 \\
		\hline\hline
	\end{tabular}
	\label{tab:T1-c-L-Mod}   
\end{table}

\section{Two benchmark models}
\label{sec:benchmark-model}

As established in Sec.~\ref{sec:intro}, a viable explanation for the observed ratio $\mathcal{B}(B^0\to K^{*0}\nu\bar{\nu})/\mathcal{B}(B^+\to K^{+}\nu\bar{\nu})$, which deviates from the SM prediction, necessitates the simultaneous generation of both $\mathcal{O}^L_V$ and $\mathcal{O}^R_V$. Our classification in Sec.~\ref{sec:classification} provides a clear pathway to achieving this. We established that diagrams ($c$) and ($d$) can individually source both operator chiralities, while diagrams ($a$)/($b$) and ($e$)/($f$) form complementary pairs that generate $\mathcal{O}^R_V$ and $\mathcal{O}^L_V$, respectively. While a simple combination of any two such valid models could suffice, this approach is often unappealing as it often introduces a large number of new particles. We therefore impose an additional criterion of minimality, seeking the most parsimonious models that can accomplish the task. 
\begin{table}[t!]
	\centering
	\caption{Particle content and SM gauge representations for the scalar-rich (left) and fermion-rich (right) benchmark models.}
	\label{tab:benchmark_models}
	\renewcommand\arraystretch{1.5}
	\tabcolsep=0.4cm 
	\begin{tabular}{ccccccccccc}
		\hline\hline
		& \multicolumn{5}{c}{Scalar-Rich Model} && \multicolumn{4}{c}{Fermion-Rich Model} \\
		\cline{2-6} \cline{8-11}
		& $\Psi$ & $S_1$ & $S_2$ & $S_3$ & $S_4$ && $\Psi_1$ & $S$ & $\Psi_2$ & $\Psi_3$ \\
		\hline
		SU(3)$_c$ & 1 & $\bar{3}$ & 1 & $\bar{3}$ & 1 && 3 & 1 & 1 & 3 \\
		SU(2)$_L$ & 1 & 1 & 2 & 2 & 2 && 1 & 1 & 2 & 2 \\
		U(1)$_Y$ & $\alpha$ & $\alpha+\frac{1}{3}$ & $\alpha-\frac{1}{2}$ & $\alpha-\frac{1}{6}$ & $\alpha+\frac{1}{2}$ && $\alpha$ & $\alpha+\frac{1}{3}$ & $\alpha-\frac{1}{6}$ & $\alpha+\frac{1}{2}$ \\
		\hline\hline
	\end{tabular}
\end{table}
  
By systematically surveying all valid model combinations, we have identified two particularly economical scenarios. The most minimal model, containing only four new fields, arises from combining two distinct solutions derived from diagram ($c$). A second competitive scenario, containing five new fields, can be constructed by combining a model from diagram ($a$) with one from diagram ($e$). All other combinations result in models with six or more new degrees of freedom and are not considered further in this work.

Based on this analysis, we select two benchmark models for detailed study, whose particle content and SM gauge representations are defined in Table~\ref{tab:benchmark_models}. The first, which we term the ``scalar-rich model'', is notable for its primarily scalar content. 
Interestingly, for the specific hypercharge assignment $\alpha=0$, the states $S_1$ and $S_3$ in this model correspond precisely to the two scalar leptoquarks introduced in Ref.~\cite{Bause:2023mfe} to address the anomalies via tree-level exchanges. 
The second, termed the ``fermion-rich model'', is instead dominated by new fermionic degrees of freedom. 

We first consider the scalar-rich model. This benchmark model is constructed from a combination of diagram ($a$), which generates $\mathcal{O}^R_V$, and diagram ($e$), which generates $\mathcal{O}^L_V$. To make the model minimal, we identify the internal fermion in both diagrams as a single field, $\Psi$, while the four scalars are treated as distinct fields, relabeled $S_{1,\dots,4}$. The relevant interaction Lagrangian is
\begin{align}\label{eq:scalar-rich}
	\mathcal{L}&\supset g^i_1\bar{\Psi} P_R d_i S_1+g^i_2\bar{\Psi}P_R L^{c}_i S_2 
	+g^i_{1^\prime}\bar{\Psi} P_L Q_iS_3+ g^i_{2^\prime}\bar{\Psi}P_L L_{i}S_4+\text{h.c.} \nonumber \\
	&=g^i_1\bar{\Psi} P_R d_i S_1+g^i_2\bar{\Psi}P_R L^{c}_{i\alpha} S_{2\alpha} 
	+g^i_{1^\prime}\bar{\Psi} P_L Q_{i\alpha}S_{3\beta}\epsilon_{\alpha\beta}+ g^i_{2^\prime}\bar{\Psi}P_L L_{i\alpha}S_{4\beta}\epsilon_{\alpha\beta}+\text{h.c.}
\end{align}
where $i$ is a generation index, $L^c_i$ represents the charge conjugate of the lepton doublet $L_i$, and $\alpha, \beta$ are $SU(2)_L$ indices contracted trivially (diagonally) or via the totally antisymmetric tensor $\epsilon_{\alpha\beta}$.

\begin{figure}[t!]
	\centering
	\includegraphics[width=0.95\textwidth]{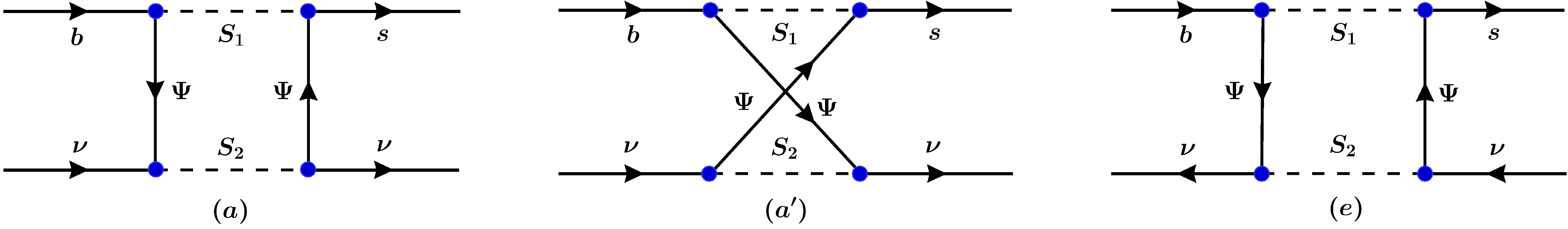}
	\caption{Illustration of the new diagrammatic contribution that becomes possible in the scalar-rich model when the mediating $\Psi$ is a Majorana particle. The box diagram ($a$) is now accompanied by a ``crossed'' diagram, denoted ($a^\prime$), where the internal fermion lines are exchanged. Note that this crossed diagram is topologically equivalent to the diagram ($e$).}
	\label{fig:cross01} 
\end{figure}

This model exhibits a special behavior when the hypercharge parameter $\alpha=0$, which allows $\Psi$ to be a Majorana fermion. This possibility enables new ``crossed'' diagrams. As shown in Fig.~\ref{fig:cross01}, crossing the internal fermion lines of diagram ($a$) produces a new diagram ($a^\prime$), which is topologically identical to diagram ($e$). Similarly, the crossed version of diagram ($e$) is equivalent to the diagram ($a$) in topology. Consequently, the Majorana nature of $\Psi$ provides an additional mechanism for generating the required operators from the same fundamental Lagrangian.

Integrating out the new degrees of freedom yields the effective Lagrangian, 
\begin{align}\label{eq:L_SR}
	\mathcal{L}_{\text{eff}}&=[C^{(1)}_{\ell q}]_{mnji}Q^{(1)}_{\ell q}+[C_{\ell d}]_{mnji}Q_{\ell d}\,,
\end{align} 
with the Wilson coefficients given by 
\begin{align}
	[C_{\ell d}]_{mnji}=&-\frac{(g^i_1g^{j*}_1g^m_2g^{n*}_2)}{4}\left[J_4(m_{S_1},\!m_{S_2},\!m_{\Psi},\!m_{\Psi})+
	2\eta^M m_{\Psi}^2 I_4(m_{S_1},\!m_{S_2},\!m_{\Psi},\!m_{\Psi}) \right]\nonumber \\
	&-\frac{(g^i_1g^{j*}_1g^n_{2^\prime}g^{m*}_{2^\prime})}{4}
	\left[2 m_{\Psi}^2 I_4(m_{S_1},\!m_{S_4},\!m_{\Psi},\!m_{\Psi})+
	\eta^MJ_4(m_{S_1},\!m_{S_4},\!m_{\Psi},\!m_{\Psi}) \right]\,,	\label{eq:SRCld}						\\		
	[C^{(1)}_{\ell q}]_{mnji}=&\frac{(g^i_{1^\prime}g^{j*}_{1^\prime}g^n_{2^\prime}g^{m*}_{2^\prime})}{4}\left[J_4(m_{S_3},\!m_{S_4},\!m_{\Psi},\!m_{\Psi})+
	2\eta^M m_{\Psi}^2 I_4(m_{S_3},\!m_{S_4},\!m_{\Psi},\!m_{\Psi}) \right]\nonumber \\
	&+\frac{g^i_{1^\prime}g^{j*}_{1^\prime}g^m_2g^{n*}_2}{4}\left[2 m_{\Psi}^2 I_4(m_{S_2},\!m_{S_3},\!m_{\Psi},\!m_{\Psi})+
	\eta^MJ_4(m_{S_2},\!m_{S_3},\!m_{\Psi},\!m_{\Psi}) \right]\,. \label{eq:SRClq}
\end{align} 
Here, the factor $\eta^M=1$ parameterizes the additional contributions from a Majorana fermion $\Psi$ (for which $\eta^M=0$, they are absent), and $I_4$ is a corresponding new loop function defined in Appendix~\ref{app:loopfunc}. 

\begin{figure}[t!]
	\centering
	\includegraphics[width=0.95\textwidth]{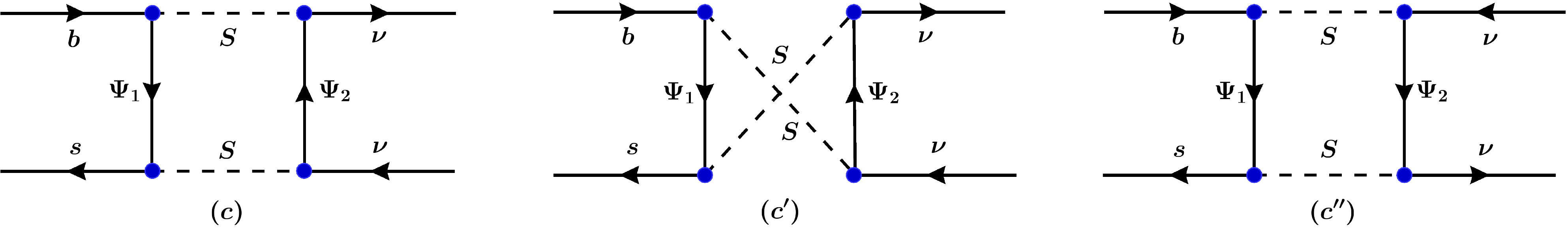}
	\caption{Illustration of additional diagrammatic contributions in the fermion-rich model that arise when the scalar mediator $S$ is a real field. The box diagram ($c$) is accompanied by a ``crossed-scalar'' diagram ($c^\prime$). This crossed diagram can be topologically rearranged to show its equivalence with diagram ($c^{\prime\prime}$), which is itself identical to the box diagram ($d$).}
	\label{fig:cross02} 
\end{figure} 

We now move on to the fermion-rich model, which is constructed by unifying two distinct solutions from diagram ($c$) to generate both $\mathcal{O}^L_V$ and $\mathcal{O}^R_V$. To achieve this minimal spectrum, we postulate that the internal scalar is a single field, $S$, common to both solutions, and that one of the internal fermions is also common to both, which we label $\Psi_2$. The remaining two fermions are distinct fields, denoted as $\Psi_1$ and $\Psi_3$. This specific combination of fields leads to the following interaction Lagrangian:
\begin{align}\label{eq:fermion-rich}
	\mathcal{L}&=g^i_1\bar{\Psi}_1 P_R d_i S+g^i_2\bar{L}^i P_R \Psi_2 S^{\dagger}
	+g^i_{1^\prime}\bar{\Psi}_3 P_L Q_iS+\text{h.c.}\nonumber \\
	&=g^i_1\bar{\Psi}_1 P_R d_i S+g^i_2\bar{L}^i_\alpha P_R \Psi_{2\alpha} S^{\dagger}
	+g^i_{1^\prime}\bar{\Psi}_{3\alpha} P_L Q_{i\alpha}S+\text{h.c.},
\end{align}
where $i$ denotes the generation index and $\alpha$ is the $SU(2)_L$ index.

As in the scalar-rich case, the possibility of the scalar $S$ being a real field (i.e., $\alpha=-1/3$) allows for contributions from ``crossed-scalar'' diagrams. As illustrated in Fig.~\ref{fig:cross02}, the box diagram ($c$) is now accompanied by a crossed diagram ($c^\prime$), where the internal scalar lines are exchanged. This new diagram ($c^\prime$) can be topologically rearranged to show its equivalence with the diagram ($d$), via the intermediate form ($c^{\prime\prime}$). 

Integrating out the new particles in the fermion-rich model yields the same operator structure as in Eq.~\eqref{eq:L_SR}, but with Wilson coefficients that depend critically on whether the scalar $S$ is real or complex. If $S$ is real, a cancellation occurs between the standard and ``crossed-scalar'' diagrams. We parameterize this with a factor $\eta^M$, set to 1 for a real scalar and 0 for a complex one, which leads to the coefficients
\begin{align}
	[C_{\ell d}]_{mnji}&=\frac{(g^i_1g^{j*}_1 g^{m}_2 g^{n*}_2)}{4}J_4(m_{\Psi_1},\!m_{\Psi_2},\!m_{S},\!m_{S}) 
	\left(1-	\eta^M\right)\,, \label{eq:FRCld}\\
	[C^{(1)}_{\ell q}]_{mnji}&=\frac{(g^i_{1^\prime}g^{j*}_{1^\prime}g^m_2g^{n*}_{2})	}{4}J_4(m_{\Psi_2},\!m_{\Psi_3},\!m_{S},\!m_{S})
	\left(1-	\eta^M\right)\,. \label{eq:FRClq}		
\end{align} 
Evidently, both coefficients vanish for a real scalar mediator, a result consistent with the findings of Ref.~\cite{Arnan:2016cpy}.

\section{Observables and constraints on Wilson coefficients}
\label{sec:Constraints}

In the previous section, we introduced two benchmark setups, corresponding to a scalar-rich and a fermion-rich scenario. The viability of these models depends on their masses and couplings, which we analyze under the simplifying assumption that the new states couple exclusively to third-generation leptons. However, the interactions generating the desired $b \to s\nu\bar{\nu}$ operators inevitably induce other flavor-changing neutral current (FCNC) processes. These include contributions to $B_s \to \tau^+ \tau^-$, $B \to K \tau^+ \tau^-$, $B_s\text{--}\bar{B}_s$ mixing, radiative $B$ decays, and $Z$-boson decays, all of which are subject to stringent experimental 
constraints.\footnote{Note that several models in Table~II and the Appendix~\ref{app:models} also generate 
	the triplet operator $\mathcal{O}_{\ell q}^{(3)}$, which is subject to 
	$b \to c \tau \bar{\nu}_\tau$ constraints, such as measurements of the 
	$R(D^{(*)})$ ratios~\cite{HeavyFlavorAveragingGroupHFLAV:2024ctg}. 
	However, this operator is absent in our two benchmark models [Eq.~(9)].} Therefore, a complete phenomenological analysis is required to identify the viable parameter space. In this section, we first evaluate the Wilson coefficients for all relevant constraining observables and then use the current experimental bounds to constrain the masses and couplings of the new particles.

\subsection{$B \to K^{(*)}\nu\bar{\nu}$}
\label{subsec:BKnunu}

We begin by establishing the LEFT operators for $b\to s{\nu}_{\ell}{\bar{\nu}}_{\ell}$ transitions. Under the assumptions that right-handed neutrinos are absent, all neutrinos are Dirac, and lepton flavor is conserved, the most general effective Lagrangian is
\begin{align}\label{eq:Lbnunu}
	\mathcal{L}_{\text{eff}} = \frac{8G_F}{\sqrt{2}}V_{tb}V_{ts}^*\frac{\alpha}{4\pi} \sum_{\ell=e,\mu,\tau}
	\left( C_L^{\nu_\ell}\mathcal{O}_V^{L} + C_R^{\nu_\ell}\mathcal{O}_V^{R} \right) + \text{h.c.}\,,
\end{align}
where $G_F$ is the Fermi constant, $\alpha$ is the fine-structure constant, and $V_{ij}$ denote the CKM matrix elements. The operators $\mathcal{O}_V^{L,R}$ are defined in Eq.~\eqref{eq:LowOp}. In the SM, the right-handed current is absent ($C_{R,\text{SM}}^{\nu_\ell}=0$), while the left-handed Wilson coefficient is predicted to be $C_{L,\text{SM}}^{\nu_\ell} = -6.32(7)$~\cite{Buras:2014fpa,Brod:2021hsj,Bause:2023mfe}. We parameterize NP contributions as shifts to these coefficients, denoted by $\Delta C_{L,R}^{\nu_\ell}$.

To connect our high-scale model to this framework, we match the SMEFT Wilson coefficients onto the LEFT basis at the electroweak scale. Using the \texttt{DsixTools 2.0} package~\cite{Celis:2017hod,Fuentes-Martin:2020zaz} to evaluate the renormalization group (RG) running effects, we find that they are approximately $3\%$ and $0.3\%$ for $\Delta C_L^{\nu_\ell}$ and $\Delta C_R^{\nu_\ell}$, respectively. Neglecting these small running corrections, we obtain
\begin{align}
	\Delta C_L^{\nu_\ell}&=\frac{\pi}{\sqrt{2}\alpha G_F V_{tb}V_{ts}^*} [C_{\ell q}^{(1)}]_{\ell\ell23} \,, \\
	\Delta C_R^{\nu_\ell}&=\frac{\pi}{\sqrt{2}\alpha G_F V_{tb}V_{ts}^*} [C_{\ell d}]_{\ell\ell23}\,.
\end{align}

The branching ratios for the decays $B \to K^{(*)}\nu\bar{\nu}$ can be expressed numerically in terms of these Wilson coefficients. Using the form factors adopted in Refs.~\cite{Buras:2014fpa, Becirevic:2023aov}, the numerical expressions for the branching fractions are given by~\cite{Chen:2024jlj}
\begin{align}
	\mathcal{B}(B^+ \to K^+\nu\bar{\nu}) &= (3.98^{+0.15}_{-0.15} \times 10^{-8} )\sum_{\ell=e,\mu,\tau} |C_L^{\nu_\ell} + C_R^{\nu_\ell}|^2\,,  \\
	\mathcal{B}(B^0 \to K^{*0}\nu\bar{\nu}) &= (6.94^{+1.72}_{-1.51} \times 10^{-8})\sum_{\ell=e,\mu,\tau} |C_L^{\nu_\ell} - C_R^{\nu_\ell}|^2	+ (1.36^{+0.23}_{-0.21} \times 10^{-8}) \sum_{\ell=e,\mu,\tau} |C_L^{\nu_\ell} + C_R^{\nu_\ell}|^2\,,
\end{align}
where the Wilson coefficients are decomposed into their SM and NP contributions, 
$C_{L(R)}^{\nu_\ell}=C_{L(R),\text{SM}}^{\nu_\ell}+\Delta C_{L(R)}^{\nu_\ell}$.
The quoted uncertainties in the numerical coefficients arise from the propagation of uncertainties in the hadronic form factors. 

The above expressions allow the experimental measurements of $B^0 \to K^{*0}\nu\bar{\nu}$ decays to be translated into constraints on the Wilson coefficients. In particular, the Belle II measurement of the charged mode yields~\cite{Belle-II:2023esi}
\begin{align}
\mathcal{B}(B^+ \to K^+\nu\bar{\nu})_\text{exp} = (2.3 \pm 0.7) \times 10^{-5},
\end{align}
while the Belle search for the neutral vector mode provides the upper limit~\cite{Belle:2017oht}
\begin{align}
\mathcal{B}(B^0 \to K^{*0}\nu\bar{\nu})_\text{exp} < 1.8 \times 10^{-5}  \quad (90\% \, \text{C.L.}).
\end{align}
These experimental results are subsequently used to constrain the allowed values of the Wilson coefficients $C_{L(R)}^{\nu_\ell}$.

\subsection{$b \to s \tau^+ \tau^-$}

The next set of constraints arises from $b \to s\tau^+\tau^-$ transitions, described by the low-energy effective Lagrangian
\begin{align}\label{eq:Lbll}
	\mathcal{L}_{\text{eff}}= \frac{4G_F}{\sqrt{2}}V_{tb}V_{ts}^* \frac{\alpha}{4\pi} \sum_{i=9,10} ( C_i \mathcal{O}_i + C_i^{\prime} \mathcal{O}_i^{\prime} ) + \text{h.c.}\,,
\end{align}
with the four-fermion operators defined as
\begin{align}
	\mathcal{O}_9^{(\prime)} = (\bar{s}\gamma_\mu P_{L(R)} b)(\bar{\tau}\gamma^\mu \tau)\,,  \qquad 
	\mathcal{O}_{10}^{(\prime)} = (\bar{s}\gamma_\mu P_{L(R)} b)(\bar{\tau}\gamma^\mu \gamma_5 \tau)\,.
\end{align}
In the SM, only the unprimed operators are generated, with Wilson coefficients evaluated at the scale $\mu_b=4.2 \text{ GeV}$ to be $C_{9,\text{SM}} \approx 4.11$ and $C_{10,\text{SM}} \approx -4.19$ at next-to-next-to-leading logarithmic order~\cite{Blake:2016olu}.

In general, NP contributions to these operators can originate from box, $Z$-penguin, and photon-penguin diagrams. However, for the class of models considered here, the box contributions dominate. Since our models do not introduce new sources of SU(2) breaking, $Z$-penguin diagrams are suppressed by factors of $m_b^2/M_Z^2$. Similarly, photon-penguin contributions to $C_9$ are strongly constrained by $b \to s\gamma$ and $B_s - \bar{B}_s$ mixing data and are subdominant in the parameter space of interest~\cite{Arnan:2016cpy}. 
Focusing therefore on the dominant box diagrams, we match the generated SMEFT operators $Q^{(1)}_{\ell q}$ and $Q_{\ell d}$ onto the LEFT basis, yielding the relations:
\begin{align}\label{eq:C9C10}
	\Delta C_9 = - \Delta C_{10} &= \frac{\pi}{\sqrt{2}\alpha G_F V_{tb}V_{ts}^*} [C_{\ell q}^{(1)}]_{3323}\,, \\ \label{eq:C9C10p}
	\Delta C_9^{\prime} = - \Delta C_{10}^{\prime} &=\frac{\pi}{\sqrt{2}\alpha G_F V_{tb}V_{ts}^*} [C_{\ell d}]_{3323}\,.
\end{align}

The effective Lagrangian in Eq.~\eqref{eq:Lbll} can mediate the rare decay $B_s \to \tau^+\tau^-$, providing a direct constraint on the Wilson coefficients $C_{10}^{(\prime)}$. The branching ratio is given by
\begin{align}
	\mathcal{B}(B_s \to \tau^+\tau^-) &= \tau_{B_s} \frac{G_F^2 \alpha^2}{16\pi^3} |V_{ts}V_{tb}^*|^2 f_{B_s}^2 m_{B_s} m_\tau^2 \sqrt{1 - \frac{4m_\tau^2}{m_{B_s}^2}} |C_{10} - C_{10}^{\prime}|^2\,,
\end{align}
where $C_{10}^{(\prime)}=C_{10,\text{SM}}^{(\prime)}+\Delta C_{10}^{(\prime)}$. Note that the vector operators $\mathcal{O}_9^{(\prime)}$ do not contribute to this purely leptonic final state due to parity conservation in the hadronic matrix element. The current experimental constraint is an upper limit set by the LHCb collaboration, $\mathcal{B}(B_s \to \tau^+\tau^-)_{\text{exp}} < 6.8 \times 10^{-3}$ at 95\% C.L.~\cite{LHCb:2017myy}. Imposing this bound on the theoretical prediction yields the constraint:
\begin{align}
	|\Delta C_{10}-\Delta C_{10}^{\prime}-4.19|^2\leq 1.60^{+0.07}_{-0.06}\times 10^{5}\,.
\end{align}
where the uncertainty stems from the propagation of $f_{B_s}$ error. 

While this constraint is currently loose, the sensitivity is expected to improve in the coming years. Future LHCb upgrades project a limit of approximately $5\times10^{-4}$~\cite{Albrecht:2017odf}, and future colliders such as the Circular Electron Positron Collider (CEPC) could potentially reach sensitivities of order $10^{-5}$~\cite{CEPCStudyGroup:2018ghi}.

Finally, we examine the semi-leptonic decay $\bar{B} \to K \tau^+ \tau^-$. The differential decay rate with respect to the dilepton invariant mass squared $s \equiv q^2$ is given by~\cite{Bobeth:2001sq}:
\begin{align}
	\frac{d\Gamma(\bar{B} \to K \tau^+ \tau^-)}{ds} &= 2 \Gamma_0 \lambda^{1/2}(M_B^2,M_K^2,s)\beta_\tau \biggl\{ s |F_P|^2+ \frac{1}{6}\lambda(M_B^2,M_K^2,s)\left(1 + \frac{2m_\tau^2}{s}\right)\Bigl(|F_A|^2 + |F_V|^2 \Bigr) \nonumber \\
	&\quad + 4m_\tau^2 M_B^2 |F_A|^2  + 2m_\tau(M_B^2 - M_K^2 + s)\mathrm{Re}\left(F_P F_A^* \right) \biggr\},
\end{align}
where the kinematic range is $4m_\tau^2 \leq s \leq (M_B - M_K)^2$. The normalization factor and kinematic functions are defined as
\begin{equation}
	\Gamma_0 = \frac{G_F^2\alpha^2}{2^9\pi^5 M_B^3}|V_{tb}V_{ts}^*|^2, \quad
	\beta_\tau = \sqrt{1 - 4m_\tau^2/s}\,, \quad 
	\lambda(a,b,c) = a^2 + b^2 + c^2 - 2(ab + bc + ac)\,.
\end{equation}
The dynamical functions $F_{P,A,V}$ incorporate contributions from both unprimed and primed operators:
\begin{align}
	F_P &= -m_\tau  (C_{10}+C_{10}^\prime)  \left[f_+(s) - \frac{M_B^2 - M_K^2}{s}\left(f_0(s) - f_+(s)\right)\right]\,, \nonumber \\
	F_A &= (C_{10}+C_{10}^\prime) f_+(s)\,, \nonumber\\
	F_V &= (C_{9}+C_{9}^\prime) f_+(s)\,.
\end{align}
Adopting the vector ($f_+$) and scalar ($f_0$) form factors from Refs.~\cite{Buras:2014fpa,Becirevic:2023aov} and applying the BaBar upper limit $\mathcal{B}(B^+ \to K^+ \tau^+\tau^-) \leq 2.25 \times 10^{-3}$ (90\% C.L.)~\cite{BaBar:2016wgb}, we obtain the following numerical constraint on the NP contributions (employing the relations in Eqs.~\eqref{eq:C9C10} and \eqref{eq:C9C10p}):
\begin{align}
 	 17.6^{+0.45}_{-0.44} 
 	+ 1.02^{+0.03}_{-0.03} | \Delta C_9 +\Delta C^{\prime}_9 |^2 + 8.46^{+0.21}_{-0.21}  \mathrm{Re}( \Delta C_9 +\Delta C^{\prime}_9)  \leq 2.25 \times 10^{5}\,.
\end{align}
 
\subsection{$B_s-\bar{B}_s$ mixing}

\begin{figure}[t!]
	\centering
	\includegraphics[width=0.7\textwidth]{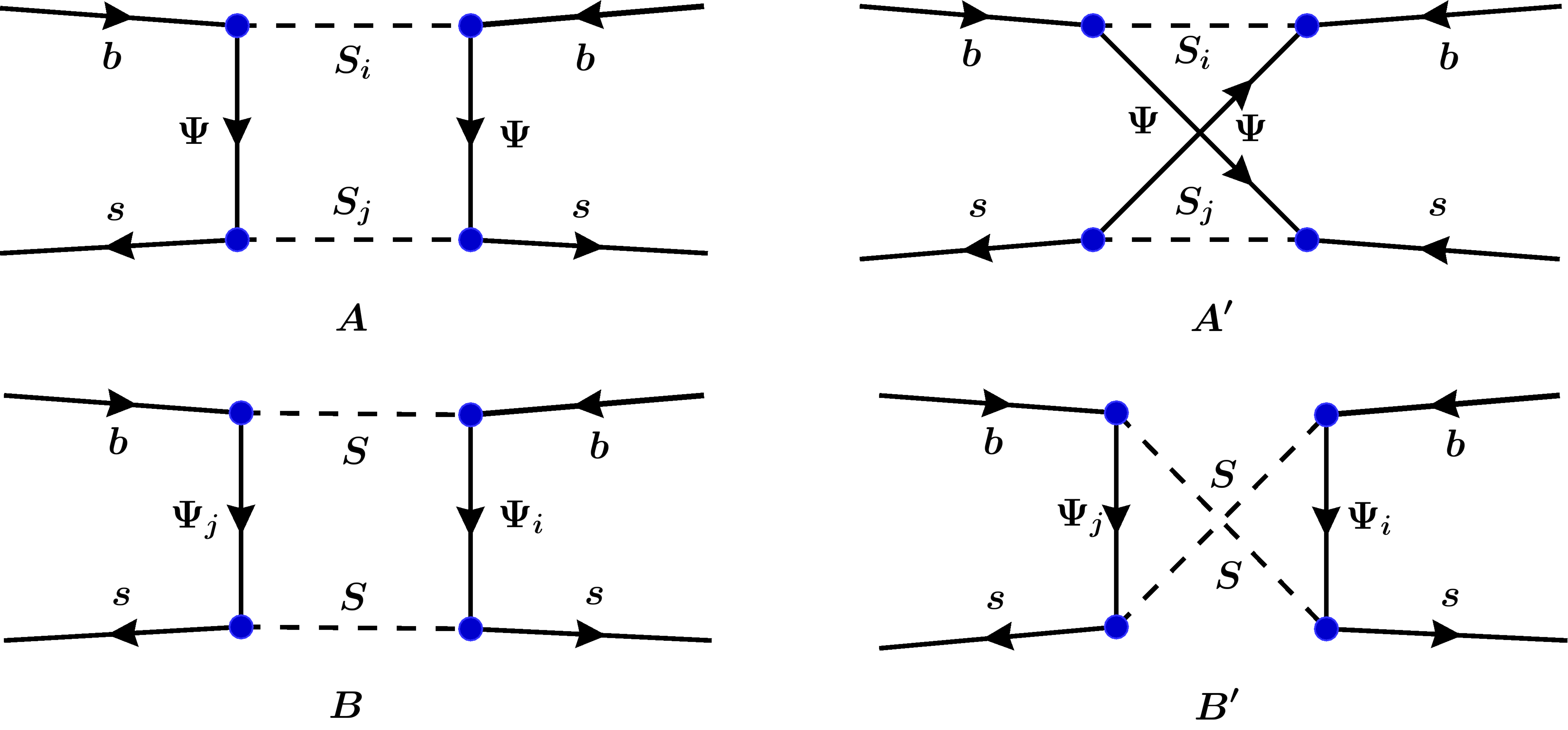}
	\caption{One-loop diagrams contributing to $B_s-\bar{B}_s$ mixing. The top panel shows diagrams for the scalar-rich model, where the standard box (A) is accompanied by a ``crossed-fermion'' diagram (A$^\prime$) if the mediator $\Psi$ is a Majorana particle. The bottom panel shows diagrams for the fermion-rich model, where the standard box (B) is accompanied by a ``crossed-scalar'' diagram (B$^\prime$) if the mediator $S$ is a real scalar. The indices $i,j$ on the mediators correspond to the distinct particle species defined in each model.}
	\label{fig:BBbar} 
\end{figure} 

Another set of constraints on our models arises from their potential contributions to $B_s-\bar{B}_s$ mixing. At low energies, this process is described by the effective Hamiltonian
\begin{align}\label{eq:Hbbbar}
	\mathcal{H}_{\text{eff}}=\sum_a C_a \mathcal{O}_a+ \text{h.c.}
\end{align} 
Adopting the so-called BMU basis~\cite{Buras:2000if}, the relevant operators for our analysis are listed as follows:
\begin{align}
	\mathcal{O}_{\text{VLL}} &= (\bar{s}\gamma_\mu P_L b)(\bar{s}\gamma^\mu P_L b)\,, \quad
	\mathcal{O}_{\text{VRR}} = (\bar{s}\gamma_\mu P_R b)(\bar{s}\gamma^\mu P_R b)\,, \quad
	\mathcal{O}_{\text{LR,1}} = (\bar{s}\gamma_\mu P_L b)(\bar{s}\gamma^\mu P_R b)\,.
\end{align}
While the SM contribution to this process is dominated by $\mathcal{O}_{\text{VLL}}$, its effect is CKM-suppressed and can be safely neglected~\cite{Aebischer:2020dsw}.
In the SMEFT framework, our benchmark models generate contributions to three of the primary $\Delta B=2$ operators, namely
\begin{align}\label{eq:bbbar}
	Q^{(1)}_{qq}&=\left(\bar{Q}_{2}\gamma^{\mu}Q_{3}\right)\left(\bar{Q}_{2}\gamma^{\mu}Q_{3}\right)\,, \\
	Q^{(1)}_{qd}&=\left(\bar{Q}_{2}\gamma^{\mu} Q_{3}\right)\left(\bar{d}_{R,2}\gamma^{\mu}d_{R,3}\right)\,, \\
	Q_{dd}&=\left(\bar{d}_{R,2}\gamma^{\mu} d_{R,3}\right)\left(\bar{d}_{R,2}\gamma^{\mu}d_{R,3}\right)\,,
\end{align}
where the subscript denotes the generation of quark fields. 
Evaluating the SMEFT-to-LEFT matching and RG evolution down to $\mu = m_b$ with \texttt{DsixTools}~\cite{Celis:2017hod,Fuentes-Martin:2020zaz} yields
\begin{equation}\label{eq:relations}
	\begin{aligned}
		C_{\mathrm{VLL}}(m_b) &\approx 0.80 \left[ C^{(1)}_{qq} \right]_{2323} , \\
		C_{\mathrm{VRR}}(m_b) &\approx 0.77 \left[ C_{dd} \right]_{2323} , \\
		C_{\mathrm{LR},1}(m_b) &\approx \left[ C^{(1)}_{qd} \right]_{2323} ,
	\end{aligned}
\end{equation}
where an approximately 5\% enhancement in $C_{\mathrm{LR},1}(m_b)$ from RG running has been neglected.

The one-loop box diagrams that generate these SMEFT operators in our benchmark models are depicted in Fig.~\ref{fig:BBbar}. As in the previous sections, the contributions depend on whether the mediating particles are complex or real. For the scalar-rich model, where the fermion $\Psi$ can be a Majorana particle ($\eta^M=1$), we obtain
\begin{align}
	[C^{(1)}_{qd}]_{2323}&=\frac{(g^3_1g^{2*}_1g^3_{1^\prime}g^{2*}_{1^\prime})}{4}\Big[2m_{\Psi}^2I_4(m_{S_1},m_{S_3},m_{\Psi},m_{\Psi})
	+\eta^M J_4(m_{S_1},m_{S_3},m_{\Psi},m_{\Psi})\Big]\,, \\
	[C_{dd}]_{2323}&=-\frac{(g^3_1g^{2*}_1)^2}{8}\Big[J_4(m_{S_1},m_{S_1},m_{\Psi},m_{\Psi})+2\eta^M m_{\Psi}^2I_4(m_{S_1},m_{S_1},m_{\Psi},m_{\Psi})\Big]\,, \\
	[C^{(1)}_{qq}]_{2323}&=-\frac{(g^3_{1^\prime}g^{2*}_{1^\prime})^2}{8}\Big[J_4(m_{S_3},m_{S_3},m_{\Psi},m_{\Psi})+2\eta^M m_{\Psi}^2I_4(m_{S_3},m_{S_3},m_{\Psi},m_{\Psi})\Big]\,.
\end{align}
In the fermion-rich model, where the scalar $S$ can be real ($\eta^M=1$), we find a cancellation between the standard and crossed-box diagrams, 
leading to
\begin{align}
	[C_{dd}]_{2323}&=-\frac{(g^3_1g^{2*}_1)^2}{8} J_4(m_{S},m_{S},m_{\Psi_1},m_{\Psi_1})(1-\eta^M)\,, \\
	[C^{(1)}_{qq}]_{2323}&=-\frac{(g^3_{1^\prime}g^{2*}_{1^\prime})^2}{8}J_4(m_{S},m_{S},m_{\Psi_3},m_{\Psi_3})(1-\eta^M)\,, \\
	[C^{(1)}_{qd}]_{2323}&=-\frac{(g^3_1g^{2*}_1g^3_{1^\prime}g^{2*}_{1^\prime})}{4}J_4(m_{S},m_{S},m_{\Psi_1},m_{\Psi_3})(1-\eta^M)\,.
\end{align}
In both sets of equations, $\eta^M=0$ if the particle is complex (Dirac fermion or complex scalar) and $\eta^M=1$ otherwise.

The theoretical prediction for the $B_s-\bar{B}_s$ mass difference, $\Delta M_s$, is determined by the off-diagonal element of the mass matrix, 
$\Delta M_s = 2|M_{12}|$. 
This matrix element is calculated from the effective Hamiltonian in Eq.~\eqref{eq:Hbbbar} via
\begin{align}\label{eq:M12}
	M_{12} = \frac{\langle B_s|\mathcal{H}_{\text{eff}}|\bar{B}_s\rangle}{2M_{B_s}} = \frac{1}{2M_{B_s}}\sum_a C_a(\mu) \langle \mathcal{O}_a\rangle(\mu)\,,
\end{align}
where $\langle \mathcal{O}_a\rangle(\mu) \equiv \langle B_s|\mathcal{O}_a|\bar{B}_s\rangle(\mu)$ denote the hadronic matrix elements of the operators evaluated at the scale $\mu$. Although both the Wilson coefficients $C_a(\mu)$ and the matrix elements $\langle \mathcal{O}_a\rangle(\mu)$ are dependent on the renormalization scale $\mu$, the physical observable $\Delta M_s$ is scale-independent.

We confront these predictions with the precise measurement from the LHCb collaboration, $\Delta M_s =(1.1693\pm 0.0004) \times 10^{-8} \text{ MeV}$~\cite{Heinicke:2021rey}. For the theoretical inputs, we adopt the matrix elements calculated in the $\overline{\text{MS}}$-NDR scheme at $\mu=4.16$ GeV~\cite{Aebischer:2020dsw}:
\begin{align}
	\langle \mathcal{O}_{\text{VLL}} \rangle = \langle \mathcal{O}_{\text{VRR}}\rangle = 0.86\pm 0.03 \text{ GeV}^{4}\,, \quad  
	\langle \mathcal{O}_{\text{LR,1}} \rangle = -1.48\pm 0.06 \text{ GeV}^{4}\,.
\end{align}
Inserting these values into Eq.~\eqref{eq:M12}, we derive the following constraint on the NP Wilson coefficients:
\begin{align}
	\left| 0.86^{+0.03}_{-0.03} (C_{\text{VLL}}+C_{\text{VRR}}) - 1.48^{+0.06}_{-0.06} C_{\text{LR,1}} \right| \leq 6.2775 \times 10^{-11}\text{ GeV}^{-2}\,.
\end{align}  
This inequality, combined with the matching conditions in Eq.~\eqref{eq:relations}, directly constrains the parameter space of our benchmark models.

\subsection{$b \to s\gamma$}
\label{sec:bsgamma}
\begin{figure}[t!]
	\centering
	\includegraphics[width=0.8\textwidth]{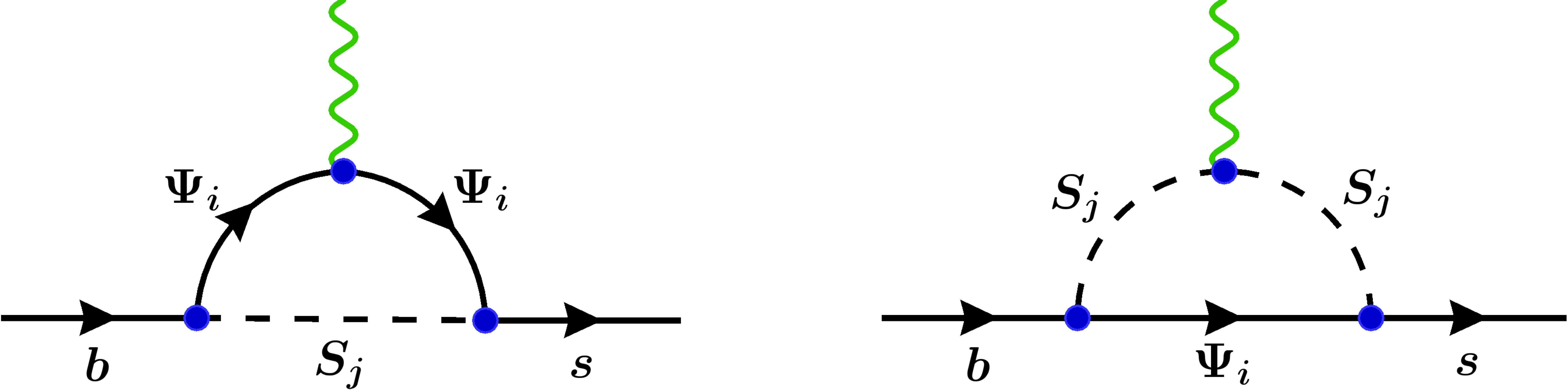}
\caption{One-loop contributions to $b\rightarrow s \gamma$ in the benchmark models. Replacing the photon (wavy green line) with a gluon yields the corresponding diagrams for the chromomagnetic transition $b\rightarrow s g$.}
	\label{fig:BSgamma} 
\end{figure}

The radiative dipole transition $b \to s\gamma$ also provides stringent constraints on our models. 
This process is described by the effective Hamiltonian
\begin{equation}
	\mathcal{H}_{\text{eff}} = -\frac{4G_F}{\sqrt{2}}V_{tb}V_{ts}^*\sum_{i=7,8} ( C_i \mathcal{O}_i + C_i^{\prime} \mathcal{O}_i^{\prime} ) \,,
\end{equation}
which contains the electromagnetic and chromomagnetic dipole operators:
\begin{equation}
	\begin{aligned}
		\mathcal{O}_7 &= \frac{e}{16\pi^2}m_b\bar{s}\sigma^{\mu\nu}P_R b F_{\mu\nu}\,, \qquad \mathcal{O}_8 = \frac{g_s}{16\pi^2}m_b\bar{s}_\alpha\sigma^{\mu\nu}P_R T^a_{\alpha\beta}b_\beta G^a_{\mu\nu}\,, \\
		\mathcal{O}^\prime_7 &= \frac{e}{16\pi^2}m_b\bar{s}\sigma^{\mu\nu}P_L b F_{\mu\nu}\,, \qquad \mathcal{O}^\prime_8 = \frac{g_s}{16\pi^2}m_b\bar{s}_\alpha\sigma^{\mu\nu}P_L T^a_{\alpha\beta}b_\beta G^a_{\mu\nu}\,.
	\end{aligned}
\end{equation}
Here, $F_{\mu\nu}$ and $G_{\mu\nu}^a$ are the electromagnetic and gluonic field strength tensors, respectively. The process $b \to s\gamma$ is directly governed by the Wilson coefficients $C_7^{(\prime)}$, which also receive contributions from the QCD mixing of the $C_8^{(\prime)}$ coefficients.

In our benchmark models, these operators are generated at the one-loop level through the diagrams shown in Fig.~\ref{fig:BSgamma}. We now calculate the new physics contributions to the Wilson coefficients, denoted as $\Delta C_{7,8}^{(\prime)}$, for each of our two models in turn.

First, we consider the scalar-rich model. In this scenario, the mediating fermion is always the color-singlet $\Psi$, while the internal scalars can be either $S_1$ or $S_3$, both of which are color anti-triplets. Integrating out these heavy particles yields the following NP contributions to the electromagnetic dipole coefficients:
\begin{align}
	\Delta C_7 &= -\frac{g_1^3 g_1^{2*}}{2m_{S_1}^2} \frac{m_s}{m_b}
	\Big(Q_{S_1} \tilde{F}_7(x_1) + Q_\Psi F_7(x_1)\Big)
	-\frac{g_{1'}^3 g_{1'}^{2*}}{2m_{S_3}^2} 
	\Big(Q_{S_3} \tilde{F}_7(x_3) + Q_\Psi F_7(x_3)\Big)\,, \\
	\Delta C_7' &= -\frac{g_1^3 g_1^{2*}}{2m_{S_1}^2}
	\Big(Q_{S_1} \tilde{F}_7(x_1) + Q_\Psi F_7(x_1)\Big) 
	-\frac{g_{1'}^3 g_{1'}^{2*}}{2m_{S_3}^2} \frac{m_s}{m_b}
	\Big(Q_{S_3} \tilde{F}_7(x_3) + Q_\Psi F_7(x_3)\Big) \,,
\end{align}
where the mass ratios are defined as $x_j = m_\Psi^2/m_{S_j}^2$ for $j=1,3$, 
and the loop functions $F_7$ and $\tilde{F}_7$ are given in Appendix~\ref{app:loopfunc}. 
The contributions to the chromomagnetic coefficients, $\Delta C_8^{(\prime)}$, are obtained by replacing the electric charges $Q_X$ with color factors $\chi_X$. In this case, $\chi_\Psi=0$ for the color singlet $\Psi$, while $\chi_{S_{1,3}}=1$ for the color anti-triplets $S_{1,3}$. 

Next, we analyze the fermion-rich model. Here, the mediating scalar $S$ is a color singlet, while the internal fermions can be either $\Psi_1$ or $\Psi_3$, both color triplets. The resulting NP contributions are
\begin{align}
	\Delta C_7 &= -\frac{g_1^3 g_1^{2*}}{2m_{S}^2} \frac{m_s}{m_b}
	\Big(Q_{S} \tilde{F}_7(x_1) + Q_{\Psi_1} F_7(x_1)\Big)
	-\frac{g_{1'}^3 g_{1'}^{2*}}{2m_{S}^2} 
	\Big(Q_{S} \tilde{F}_7(x_3) + Q_{\Psi_3} F_7(x_3)\Big)\,, \\
	\Delta C_7' &= -\frac{g_1^3 g_1^{2*}}{2m_{S}^2}
	\Big(Q_{S} \tilde{F}_7(x_1) + Q_{\Psi_1} F_7(x_1)\Big) 
	-\frac{g_{1'}^3 g_{1'}^{2*}}{2m_{S}^2} \frac{m_s}{m_b}
	\Big(Q_{S} \tilde{F}_7(x_3) + Q_{\Psi_3} F_7(x_3)) \,,
\end{align}
where the mass ratios are now defined as $x_j = m_{\Psi_j}^2/m_{S}^2$. The corresponding chromomagnetic coefficients $\Delta C_8^{(\prime)}$ are again found by substituting electric charges with color factors. Specifically, $\chi_S=0$ for the color scalar $S$ and $\chi_{\Psi_{1,3}}=1$ for the color triplets $\Psi_{1,3}$.

The theoretical prediction for the branching ratio, including the photon energy cut $E_\gamma > E_0 = 1.6\,\text{GeV}$, can be separated into the SM prediction and the NP correction:
\begin{align}
	\overline{\mathcal{B}}(b \to s\gamma) = \overline{\mathcal{B}}(b \to s\gamma)^{\text{SM}} + \delta\overline{\mathcal{B}}(b \to s\gamma)\,.
\end{align}
The NP contribution, $\delta\overline{\mathcal{B}}(b \to s\gamma)$, evaluated at scale $\mu = 160\,\text{GeV}$ 
is given by~\cite{Lunghi:2006hc, Enomoto:2015wbn}
\begin{align}\label{eq:delta_Bsg}
	\delta\overline{\mathcal{B}}(b \to s\gamma) &= 10^{-4} \times \operatorname{Re}\bigl[ -8.100\Delta \mathcal{C}_7 - 2.509\,\Delta\mathcal{C}_8+ 2.767\left(\Delta \mathcal{C}_7\Delta \mathcal{C}_8^{*}+\Delta \mathcal{C}_{7}^{\prime}\Delta \mathcal{C}_8^{\prime *}\right)  \nonumber \\
	&\quad + 5.348\left( |\Delta\mathcal{C}_7|^2+|\Delta\mathcal{C}_7^{\prime}|^2\right) + 0.890\left( |\Delta\mathcal{C}_8|^2+|\Delta\mathcal{C}_8^{\prime}|^2\right)\bigr]\,.
\end{align}
For our analysis, we have only taken account of the contributions from the leading order of $\Delta\mathcal{C}_{7,8}^{(\prime)}$. 
For the complete next-to-leading order expression, we refer the reader to the original references. 

The current world average for the experimental branching ratio is~\cite{HeavyFlavorAveragingGroupHFLAV:2024ctg}:
\begin{align}
	\mathcal{B}(B \to X_s\gamma)^{\text{exp}} &= (3.49 \pm 0.19) \times 10^{-4}\,,
\end{align}
compared to the SM prediction at the NNLO~\cite{Misiak:2015xwa}:
\begin{align}
	\mathcal{B}(B \to X_s\gamma)^{\text{SM}} &= (3.36 \pm 0.23)\times 10^{-4}\,.
\end{align}
The difference between the experimental value and the SM prediction provides a window for new physics. By requiring our calculated NP contribution, $\delta\overline{\mathcal{B}}(b \to s\gamma)$, to be consistent with this window, we can place a direct constraint on the NP Wilson coefficients $\Delta C_{7,8}^{(\prime)}$. 

\subsection{Impact on $Z\bar{f}f$ Couplings}
\label{sec:Zcoupling_analysis}

\begin{figure}[tbp]
	\centering
	\includegraphics[width=0.95\textwidth]{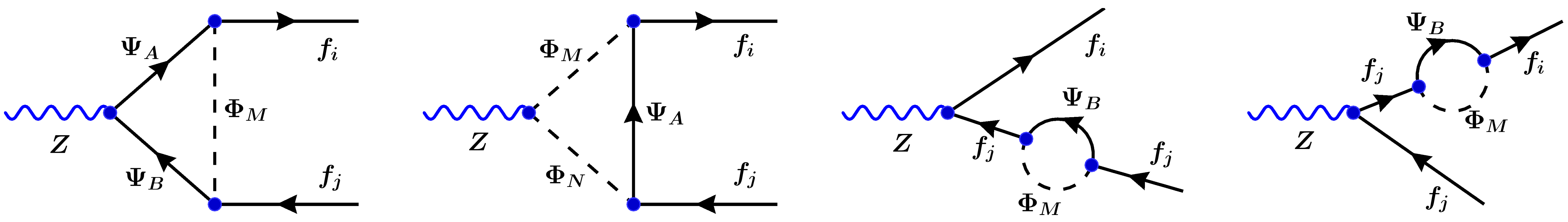}
	\caption{Feynman diagrams modifying the $Z \bar{f}_i f_j $ vertex with $f_{i,j} \in \{s, b, \ell, \nu \} $. }
	\label{fig:Zff}
\end{figure}

Constraints also arise from radiative corrections to the effective $Z$-boson couplings to SM fermions. We consider contributions affecting both on-shell $Z$ decay vertices and off-shell $Z$ exchange amplitudes. With the detailed analytical expressions for the generic one-loop functions provided in Appendix~\ref{app:Zcoupling}, we focus here on applying these results to the specific interaction structures of our two benchmark models.

For the scalar-rich model, the fermion sector contains a single SU(2)$_L$ singlet, vector-like fermion $\Psi$. Its couplings to the $Z$ boson, following the notation of Eq.~\eqref{eq:Lz}, are given by:
\begin{align}
	g^{\Psi,L}= g^{\Psi,R}=-\alpha s_W^2\,.
\end{align} 
Regarding the scalar fields, we organize them into multiplets $\Phi_M$ to match the generic formalism. We explicitly label their components by their electric charges (superscripts) and group them as follows:
\begin{align}\label{eq:phis}
	\Phi_1=\begin{pmatrix}
		S^{\alpha+\frac{1}{3}}_1 \\
		S^{\alpha+\frac{1}{3}}_3
	\end{pmatrix}\,, \quad 
	\Phi_2=\begin{pmatrix}
		S^{\alpha}_2 \\
		S^{\alpha}_4
	\end{pmatrix}\,, \quad 
	\Phi_3=S^{\alpha-1}_2\,,	\quad
	\Phi_4=S^{\alpha+1}_4\,.
\end{align} 
The components of $\Phi_1$ couple to down-type quarks $d_i$, and those of $\Phi_2$ to left-handed neutrinos $\nu_i$. In addition, $\Phi_4$ couples to 
the charged leptons $e_i$, while $\Phi_3$ interacts with the corresponding charge-conjugated fields $e^{c}_{i}$.
We omit the component $S^{\alpha-2/3}_3$, as it couples exclusively to up-type quarks. The corresponding $Z$-coupling matrices $g^{\Phi_M}$ (cf.\ Eq.~\eqref{eq:Lz}) are given by
\begin{equation}
	\begin{aligned}
		g^{\Phi_1}&=\begin{pmatrix}
			-Q_{\Phi_1} s_W^2 & 0 \\
			0 & \frac{1}{2}-Q_{\Phi_1} s_W^2
		\end{pmatrix}\,, \quad 	&g^{\Phi_3}&=-\frac{1}{2}-Q_{\Phi_3} s_W^2 \,,   \\
		g^{\Phi_2}&=\begin{pmatrix}
			\frac{1}{2}-Q_{\Phi_2} s_W^2 & 0 \\
			0 & -\frac{1}{2}-Q_{\Phi_2} s_W^2
		\end{pmatrix}\,, \quad 
		&g^{\Phi_4}&=\frac{1}{2}-Q_{\Phi_4} s_W^2 \,,
	\end{aligned}
\end{equation}
where the relevant electric charges are $Q_{\Phi_1}=\alpha+1/3$, $Q_{\Phi_2}=\alpha$, $Q_{\Phi_3}=\alpha-1$, and $Q_{\Phi_4}=\alpha+1$.

Turning to the fermion-rich model, the scalar sector consists of a single SU(2)$_L$ singlet field $S$, with a $Z$ boson coupling determined solely by its charge:
\begin{align}
	g^{S}= -(\alpha+\frac{1}{3}) s_W^2\,.
\end{align} 
The vector-like fermion sector contains multiple fields. Labeling the fermion components by their electric charges, we define the multiplets $F_{1,2,3}$ as:
\begin{align}\label{eq:psi}
	F_1=\begin{pmatrix}
		\Psi^{\alpha}_1 \\
		\Psi^{\alpha}_3
	\end{pmatrix}\,, \quad 
	F_2=\Psi^{\alpha+\frac{1}{3}}_2\,, \quad 
	F_3=\Psi^{\alpha-\frac{2}{3}}_2\,.
\end{align} 
The component $\Psi^{\alpha+1}_3$ is not presented, because it couples exclusively to up-type quarks. 
The corresponding $Z$ couplings of the multiplets $F_{1,2,3}$ (cf.\ Eq.~\eqref{eq:Lz}) are given by
\begin{equation}
	\begin{aligned}
		&g^{F_1,L}=g^{F_1,R}=\begin{pmatrix}
			-Q_{F_1} s_W^2 & 0 \\
			0 & -\frac{1}{2}-Q_{F_1} s_W^2
		\end{pmatrix}\,, \\
		&g^{F_2,L}=g^{F_2,R}=\frac{1}{2}-Q_{F_2} s_W^2 \,, \\
		&g^{F_3,L}=g^{F_3,R}=-\frac{1}{2}-Q_{F_3} s_W^2 \,,
	\end{aligned}
\end{equation}
where the relevant electric charges are $Q_{F_1}=\alpha$, $Q_{F_2}=\alpha+1/3$, and $Q_{F_3}=\alpha-2/3$.

With the couplings and field components established, we proceed to calculate the one-loop corrections to the decay modes $Z\to \bar{s}b$, $\tau^+\tau^-$, and $\nu_{\tau}\bar{\nu}_{\tau}$. In the scalar-rich model, the contributions to the effective couplings are found to be:
\begin{align}
	\Delta g_L^{sb}(m_Z^2) = & \frac{m^2_Z}{32\pi^2} g^{2*}_{1^\prime}g^{3}_{1^\prime} \left[
	\frac{g^{\Psi,L}}{M^2_{S_3}}x_3\tilde{G}_Z(x_3,x_3)
	-\frac{2}{3}\frac{g^{\Psi,R}}{M^2_{S_3}}\tilde{F}_Z(x_3,x_3)
	+\frac{\left(g^{\Phi_1}\right)_{22} }{3M^2_{\Psi}}\tilde{H}_Z(x_3,x_3)\right]\,, \\
	\Delta g_R^{sb}(m_Z^2) = & \frac{m^2_Z}{32\pi^2} g^{2*}_{1}g^{3}_{1} \left[
	\frac{g^{\Psi,R}}{M^2_{S_1}}x_1\tilde{G}_Z(x_1,x_1)
	-\frac{2}{3}\frac{g^{\Psi,L}}{M^2_{S_1}}\tilde{F}_Z(x_1,x_1)
	+\frac{\left(g^{\Phi_1}\right)_{11} }{3M^2_{\Psi}}\tilde{H}_Z(x_1,x_1)\right]\,,
\end{align}
for the quark modes. For the leptonic modes, we must account for the color charge of the internal NP fields, which yields an enhancement factor of $\chi_Z=3$. The resulting expressions are:
\begin{align}
	\Delta g_L^{\tau\tau}(m_Z^2) = & \frac{3m^2_Z}{32\pi^2}\Bigg\{ |g^{3}_{2^\prime}|^2 \left[
	\frac{g^{\Psi,L}}{M^2_{S_4}}x_4\tilde{G}_Z(x_4,x_4)
	-\frac{2}{3}\frac{g^{\Psi,R}}{M^2_{S_4}}\tilde{F}_Z(x_4,x_4)
	+\frac{g^{\Phi_4}}{3M^2_{\Psi}}\tilde{H}_Z(x_4,x_4)\right]\nonumber \\
	&-|g^{3}_{2}|^2\left[
	\frac{g^{\Psi,R}}{M^2_{S_2}}x_2\tilde{G}_Z(x_2,x_2)
	-\frac{2}{3}\frac{g^{\Psi,L}}{M^2_{S_2}}\tilde{F}_Z(x_2,x_2)
	+\frac{g^{\Phi_3}}{3M^2_{\Psi}}\tilde{H}_Z(x_2,x_2)\right]\Bigg\}\,, \\
	\Delta g_L^{\nu_\tau\nu_\tau}(m_Z^2) = & \frac{3m^2_Z}{32\pi^2}\Bigg\{ |g^{3}_{2^\prime}|^2 \left[
	\frac{g^{\Psi,L}}{M^2_{S_4}}x_4\tilde{G}_Z(x_4,x_4)
	-\frac{2}{3}\frac{g^{\Psi,R}}{M^2_{S_4}}\tilde{F}_Z(x_4,x_4)
	+\frac{\left(g^{\Phi_2}\right)_{22} }{3M^2_{\Psi}}\tilde{H}_Z(x_4,x_4)\right]\nonumber \\
	&-|g^{3}_{2}|^2\left[
	\frac{g^{\Psi,R}}{M^2_{S_2}}x_2\tilde{G}_Z(x_2,x_2)
	-\frac{2}{3}\frac{g^{\Psi,L}}{M^2_{S_2}}\tilde{F}_Z(x_2,x_2)
	+\frac{\left(g^{\Phi_2}\right)_{11} }{3M^2_{\Psi}}\tilde{H}_Z(x_2,x_2)\right]\Bigg\}\,, \\
	\Delta g_R^{\tau\tau}(m_Z^2) = &\, \Delta g_R^{\nu_\tau\nu_\tau}(m_Z^2) = 0\,.
\end{align}
Here, $M_{S_i}$ ($i=1,\dots,4$) and $M_{\Psi}$ denote the masses of the scalar fields $S_{i}$ and the fermion $\Psi$, respectively, and we have defined the mass ratios $x_i \equiv M^2_{\Psi}/M^2_{S_i}$.

Analogously, within the fermion-rich model, we calculate the one-loop corrections to the $Z \to \bar{s}b$, $\tau^+\tau^-$, and $\nu_\tau \bar{\nu}_\tau$ decay channels. The resulting contributions to the effective form factors are given by
\begin{align}
	\Delta g_L^{sb}(m_Z^2) = & \frac{m^2_Z}{32\pi^2} g^{2*}_{1^\prime}g^{3}_{1^\prime} \left[
	\frac{\left(g^{F_1,L}\right)_{22}}{M^2_{S}}y_3\tilde{G}_Z(y_3,y_3)
	-\frac{2}{3}\frac{\left(g^{F_1,R}\right)_{22}}{M^2_{S}}\tilde{F}_Z(y_3,y_3)
	+\frac{g^S}{3M^2_{\Psi_3}}\tilde{H}_Z(y_3,y_3)\right]\,, \\
	\Delta g_R^{sb}(m_Z^2) = & \frac{m^2_Z}{32\pi^2} g^{2*}_{1}g^{3}_{1} \left[
	\frac{\left(g^{F_1,R}\right)_{11}}{M^2_{S}}y_1\tilde{G}_Z(y_1,y_1)
	-\frac{2}{3}\frac{\left(g^{F_1,L}\right)_{11}}{M^2_{S}}\tilde{F}_Z(y_1,y_1)
	+\frac{g^S}{3M^2_{\Psi_1}}\tilde{H}_Z(y_1,y_1)\right]\,,
\end{align}
for the quark sector, and
\begin{align}
	\Delta g_L^{\tau\tau}(m_Z^2) = & \frac{3m^2_Z}{32\pi^2} |g^{3}_{2}|^2 \left[
	\frac{g^{F_3,L}}{M^2_{S}}y_2\tilde{G}_Z(y_2,y_2)
	-\frac{2}{3}\frac{g^{F_3,R}}{M^2_{S}}\tilde{F}_Z(y_2,y_2)
	+\frac{g^S}{3M^2_{\Psi_2}}\tilde{H}_Z(y_2,y_2)\right]\,, \\
	\Delta g_L^{\nu_\tau\nu_\tau}(m_Z^2) = & \frac{3m^2_Z}{32\pi^2} |g^{3}_{2}|^2 \left[
	\frac{g^{F_2,L}}{M^2_{S}}y_2\tilde{G}_Z(y_2,y_2)
	-\frac{2}{3}\frac{g^{F_2,R}}{M^2_{S}}\tilde{F}_Z(y_2,y_2)
	+\frac{g^S}{3M^2_{\Psi_2}}\tilde{H}_Z(y_2,y_2)\right]\,, \\
	\Delta g_R^{\tau\tau}(m_Z^2) = & \, \Delta g_R^{\nu_\tau\nu_\tau}(m_Z^2) = 0\,,
\end{align}
for the lepton sector. Here, $M_{S}$ and $M_{\Psi_k}$ (with $k=1,2,3$) denote the masses of the scalar field $S$ and the vector-like fermions $\Psi_k$, respectively. The mass ratios are defined as $y_k=M^2_{\Psi_k}/M^2_{S}$.

To constrain the new physics parameters, we utilize the precision measurements of $Z$-pole observables performed at LEP~\cite{ALEPH:2005ab}. Following the analysis in Refs.~\cite{Efrati:2015eaa, Arnan:2019uhr}, we adopt the following model-independent bounds on the effective leptonic form factors\footnote{We note that our sign convention for the $Z$ couplings aligns with Ref.~\cite{Efrati:2015eaa} but is opposite to that of Ref.~\cite{Arnan:2019uhr}.}:
\begin{align}
	\Delta g_{L}^{\tau\tau}(m_Z^2)  &= (0.16 \pm 0.58) \times 10^{-3}\,, \\
	\Delta g_{L}^{\nu\nu}(m_Z^2)  &= -(0.40 \pm 0.21) \times 10^{-2}\,. 
\end{align}
Regarding the $Z \to \bar{s}b$ decay, we apply the upper limit $\mathcal{B}(Z\to q\bar{q}^\prime) < 2.9 \times 10^{-3}$ at 95\% C.L., which is derived from the consistency between the measured hadronic $Z$ width and the SM prediction~\cite{OPAL:2000ufp,Kamenik:2023hvi}. As the SM contribution to this mode is negligible compared to the current experimental sensitivity~\cite{Kamenik:2023hvi}, we assume the decay is dominated by the NP contribution. This yields the constraint:
\begin{align}
	\frac{m_Z}{12\pi\Gamma_Z }\frac{g^2_2}{c_W^2}\left(|\Delta g_{L}^{sb}|^2+|\Delta g_{R}^{sb}|^2\right) < 2.9\times 10^{-3}\,,
\end{align}
where $\Gamma_Z$ is the total decay width of the $Z$ boson. These constraints are subsequently used to delimit the viable parameter space for the couplings in both benchmark models.

\section{Phenomenological Analysis}
\label{sec:Pheno}

Having established the analytical bounds from individual decay channels, in this section, we perform a combined analysis to delineate the viable parameter space. We begin by defining benchmark masses for the new particles consistent with LHC direct searches. Subsequently, we map the allowed regions for the Yukawa couplings under specific degeneracy schemes. Finally, we evaluate the maximum attainable branching ratios for $B \to K^{(*)} \nu \bar{\nu}$ permitted by the global data. 

Let us first establish specific benchmark values for the masses of the new scalars and fermions. For the fermion-rich model, following the analysis in Ref.~\cite{Arnan:2019uhr} to ensure safety from current bounds~\cite{ATLAS:2017avc,ATLAS:2017mjy}, we set the masses of the colored fermions to $M_{\Psi_1}=M_{\Psi_3}=3.15$~TeV, while assigning a moderate mass of $450$~GeV to the scalar $S$ and the fermion $\Psi_2$ ($M_{S}=M_{\Psi_2}=450$~GeV). This specific mass spectrum places the model well beyond the reach of direct searches at the LHC~\cite{Banerjee:2022xmu}. 

For the scalar-rich model, we align the mass spectrum of the color-singlet sector with the previous case, setting $M_{\Psi}=M_{S_2}=M_{S_4}=450$~GeV. This choice is justified because these fields are colorless and share similar quantum numbers with their counterparts in the fermion-rich model ($S$ and $\Psi_2$), implying comparable phenomenological constraints. The masses of the colored scalars $S_{1,3}$ are then constrained by their decay into the $\Psi$ fermion. Analyses of $36~\text{fb}^{-1}$ data at $\sqrt{s} = 13$~TeV~\cite{Racco:2015dxa,ATLAS:2017bfj} indicate that for the $\Psi$ mass of $450$~GeV, the lower bound on the colored scalar mass is approximately $1.10$~TeV.

Accordingly, we define the following two benchmark points (BPs):
\begin{itemize}
	\item \textbf{BP-I (Scalar-rich)}: $M_{\Psi}=M_{S_2}=M_{S_4}=450$~GeV, $M_{S_1}=M_{S_3}=1.10$~TeV.
	\item \textbf{BP-II (Fermion-rich)}: 
	$M_{S}=M_{\Psi_2}=450$~GeV, $M_{\Psi_1}=M_{\Psi_3}=3.15$~TeV.
\end{itemize}

\begin{figure}[!thbp]
	\centering
	\includegraphics[width=0.5\textwidth]{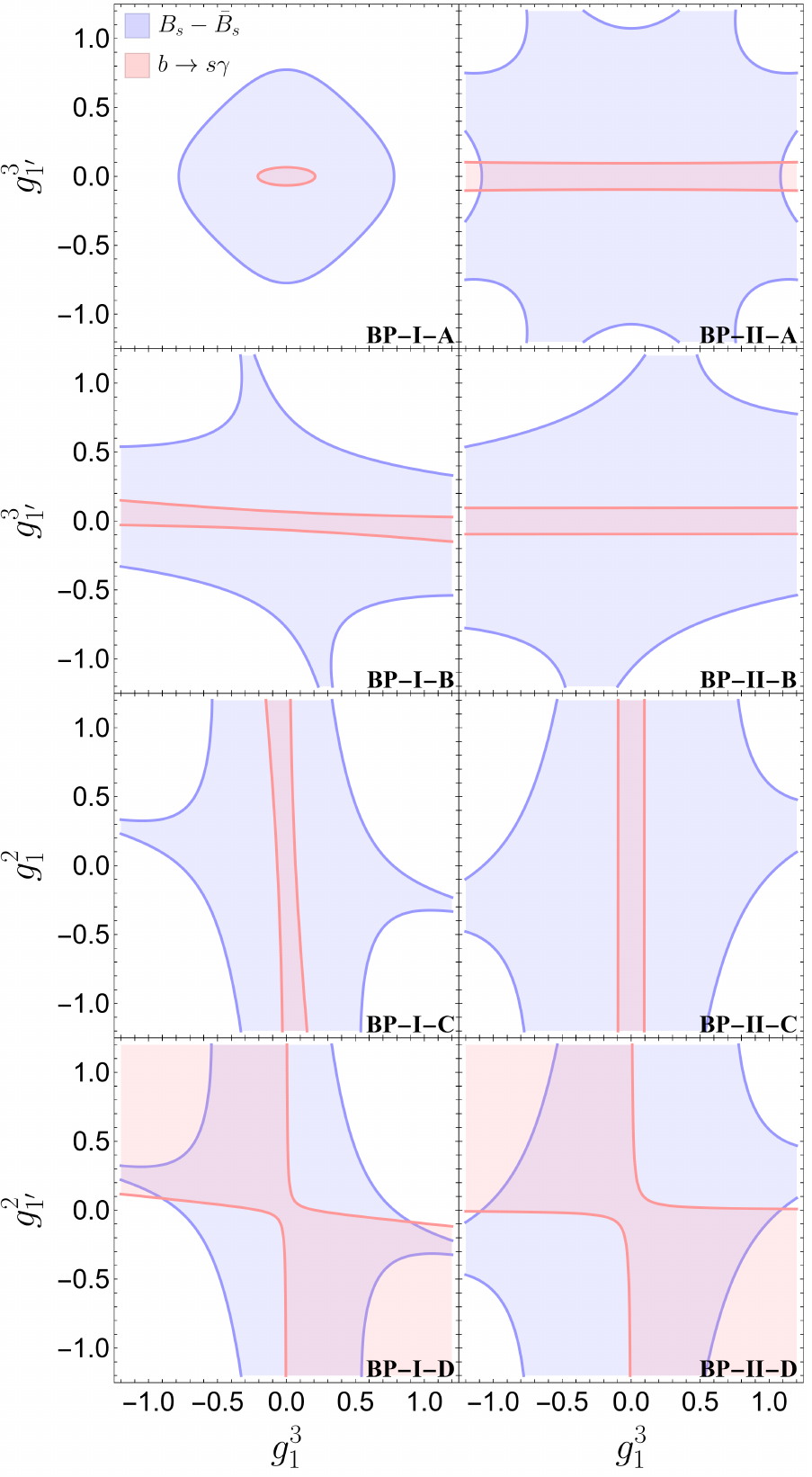}
\caption{Allowed regions for the quark couplings in the scalar-rich ($\text{BP-I}$) and fermion-rich ($\text{BP-II}$) models across the flavor structures A--D. The blue and red shaded areas depict the parameter space consistent with $B_s\text{--}\bar{B}_s$ mixing and $b \to s\gamma$ constraints, respectively. The boundaries are determined from experimental limits, with theoretical uncertainties incorporated conservatively by adopting the upper error bounds, thereby imposing more stringent constraints on the couplings. The corresponding benchmark model and flavor scheme are indicated in the bottom-right corner of each panel.}
	\label{fig:BBandbsrg}
\end{figure}

With the mass spectra established, we proceed to constrain the coupling constants. We require all dimensionless couplings to satisfy the perturbative unitarity condition, $|g| \le 1$, and assume they are real for simplicity. While the fermion-rich model is described by five couplings ($\{g^3_1, g^2_1, g^3_{1'}, g^2_{1'}, g_2^3\}$), the scalar-rich model nominally introduces a sixth, $\{g^3_1, g^2_1, g^3_{1'}, g^2_{1'}, g_2^3, g_{2'}^3\}$. To simplify the analysis, we reduce the scalar-rich parameter space by assuming $g_2^3 = g_{2'}^3$, resulting in five independent constants for both models.

We note that the constraints derived from $B_s-\bar{B}_s$ mixing, $b \to s\gamma$, and $Z \to b\bar{s}$ act exclusively on the quark-sector couplings ($g^3_1, g^2_1, g^3_{1'}, g^2_{1'}$). Similarly, the leptonic $Z$-pole observables ($Z\to \tau^+\tau^-$ and $Z\to \nu_{\tau} \bar{\nu}_{\tau}$) depend solely on the lepton coupling $g_2^3$ (and $g_{2'}^3$). We therefore utilize these processes to independently constrain the respective quark and lepton couplings. Our numerical evaluation reveals that the quark-sector couplings are strictly bounded by $B_s-\bar{B}_s$ mixing and $b \to s\gamma$. Conversely, for the lepton coupling, we find that the constraints imposed by the leptonic $Z$-pole observables are less severe than the perturbative unitarity requirement.

In contrast, the semileptonic decays ($B \to K^{(*)} \nu \bar{\nu}$ and $b \to s \tau^+\tau^-$) involve the full set of couplings, making a direct analytical inversion to constrain the parameters infeasible. We therefore treat the experimental limits on these processes as additional conditions when determining the global maximum of the branching ratios $\mathcal{B}(B^+ \to K^+ \nu \bar{\nu})$ and $\mathcal{B}(B^0 \to K^{*0} \nu \bar{\nu})$ within the allowed parameter space set by the quark and lepton sector constraints. 

 To further simplify the phenomenological analysis of the remaining five free parameters ($g^3_1, g^2_1, g^3_{1'}, g^2_{1'}, g_2^3$), we consider four distinct flavor structures for the quark couplings. In determining the allowed parameter space shown in Fig.~\ref{fig:BBandbsrg}, theoretical uncertainties are incorporated conservatively by adopting their upper error bounds, thereby imposing more stringent limits on the couplings. We have verified that varying these uncertainties within their standard ranges does not significantly alter the resulting exclusion boundaries or our qualitative conclusions. The four flavor structures are defined as follows:
\begin{itemize}
	\item \textbf{Structure A:} $ g^3_1=g^2_1, \, g^3_{1'}=g^2_{1'}$. \\
	As illustrated in the BP-I-A panel of Fig.~\ref{fig:BBandbsrg}, the constraint from $b \to s\gamma$ is generally more stringent than that from $B_s-\bar{B}_s$ mixing. Under this structure, the scalar-rich model (BP-I) is subject to significantly stronger constraints than the fermion-rich model (BP-II). Specifically, the viable parameter space for BP-I-A is restricted to a small, closed region, limiting $g^3_{1}$ to the interval $[-0.21, 0.21]$ and $g^3_{1'}$ to $[-0.07, 0.07]$. In contrast, the $b \to s\gamma$ constraint for BP-II-A manifests as a horizontal band extending across the entire perturbative range of $g^3_{1}$. Consequently, while $g^3_{1}$ remains effectively unconstrained, $g^3_{1'}$ is restricted to a narrow interval, which reaches its minimum width of $[-0.10, 0.10]$ at $g^3_{1}=0$.
	
	\item \textbf{Structure B:} $ g^2_1=g^3_{1'}=g^2_{1'} $. \\
	As depicted in the panels labeled BP-I-B and BP-II-B in Fig.~\ref{fig:BBandbsrg}, $g^3_1$ remains largely unconstrained within the perturbative limit $[-1,1]$ for both models, while the common coupling is tightly bounded by $b \to s\gamma$. Specifically, the common coupling is restricted to approximately $\pm 0.07$ for BP-I and $\pm 0.10$ for BP-II.
	
	\item \textbf{Structure C:} $ g^3_1=g^3_{1'}=g^2_{1'} $. \\
	The resulting allowed regions (shown in panels BP-I-C and BP-II-C of Fig.~\ref{fig:BBandbsrg}) effectively mirror those of structure B with the axes interchanged, a consequence of the symmetric role these couplings play in the loop functions. Accordingly, the stringent constraints now apply to the group of couplings fixed to $g^3_1$, while $g^2_1$ remains unconstrained over the range $[-1,1]$ in both models.
	
	\item \textbf{Structure D:} $ g^3_1=g^2_1=g^3_{1'} $. \\
	As shown in the BP-I-D and BP-II-D panels of Fig.~\ref{fig:BBandbsrg}, the viable parameter space is notably larger compared to the previous schemes, indicating that decoupling $g^2_{1'}$ from the other parameters provides considerable freedom to satisfy the flavor structures. We observe that the allowed regions generally favor scenarios where $g^3_1$ and $g^2_{1'}$ exhibit opposite signs.
\end{itemize}
 
 \begin{table}[htp]
 	\centering    
 	\renewcommand\arraystretch{1.3}  
 	\setlength{\tabcolsep}{16pt}
 	\caption{Maximum attainable branching fractions for $B^+\to K^+\nu\bar{\nu}$ and $B^0\to K^{*0}\nu\bar{\nu}$ in the scalar-rich ($\text{BP-I}$) and fermion-rich ($\text{BP-II}$) models. Results are presented separately for the maximization of the charged mode (top panel) and the neutral mode (bottom panel) across the four flavor structures A--D. The column ``Param.\ Values'' lists the specific configuration $[x, y, z]$ that yields the maximum, where $x\equiv(g^3_2)^2$, $y \equiv g^3_1$, and $z$ represents the remaining independent parameter defined by the scheme. In cases where the maximum coincides with the SM limit, entries containing the symbols $x$, $y$, or $z$ rather than numerical values indicate that the parameter is unconstrained; for instance, $(x, 0, 0)$ denotes that the branching fraction is independent of $x$ when the remaining couplings vanish.  The predicted branching fraction for the complementary mode evaluated at each point, along with the resulting ratio $R$, are listed in the final two columns. Quoted uncertainties originate from the hadronic form factors.}
 	\label{tab:max_charged}
 	\fontsize{9}{13}\selectfont
 	\begin{tabular}{c|c|c|c|c|c}
 		\hline \hline
 		\multirow{2}{*}{Model} &
 		\multirow{2}{*}{\shortstack{Structure}} & \multicolumn{4}{c}{$B^+\to K^+\nu\bar{\nu}$ Optimization} \\
 		\cline{3-6}
 		& & $\mathcal{B}_{\text{max}}$ & Param.\ Values $[x, y, z]$ & $\mathcal{B}(B^0\to K^*\nu\bar{\nu})$ & $R$ \\
 		\hline \hline 
 		
 		\multirow{4}{*}{BP-I} 
 		& A & $4.78^{+0.18}_{-0.18}\times10^{-6}$ & $(1, \pm 0.21, 0)$ & $9.93^{+2.33}_{-2.06}\times10^{-6}$ & $2.07^{+0.59}_{-0.50}$  \\
 		& B & $4.78^{+0.18}_{-0.18}\times10^{-6}$ & $(1, \pm 1, \pm 0.03)$ & $9.93^{+2.33}_{-2.06}\times10^{-6}$ & $2.08^{+0.59}_{-0.49}$  \\
 		& C & $4.78^{+0.18}_{-0.18}\times10^{-6}$ & $(1, \pm 0.03, \pm 1)$ & $9.93^{+2.33}_{-2.06}\times10^{-6}$ & $2.08^{+0.59}_{-0.49}$  \\
 		& D & $5.24^{+0.20}_{-0.20}\times10^{-6}$ & $(1, \pm 1, \mp 0.31)$ & $9.71^{+2.27}_{-2.00}\times10^{-6}$ & $1.85^{+0.52}_{-0.44}$ \\
 		\hline 
 		
 		\multirow{4}{*}{BP-II} 
 		& A & $4.77^{+0.18}_{-0.18}\times10^{-6}$ & $(x, 0, 0)$,$(0, y, z)$ & $9.94^{+2.34}_{-2.06}\times10^{-6}$ & $2.09^{+0.60}_{-0.50}$  \\
 		& B & $4.78^{+0.18}_{-0.18}\times10^{-6}$ & $(1, \pm 1, \mp 0.09)$ & $9.93^{+2.33}_{-2.06}\times10^{-6}$ & $2.08^{+0.59}_{-0.50}$  \\
 		& C & $4.78^{+0.18}_{-0.18}\times10^{-6}$ & $(1, \pm 0.09, \mp 1)$ & $9.93^{+2.33}_{-2.06}\times10^{-6}$ & $2.08^{+0.59}_{-0.50}$  \\
 		& D & $4.80^{+0.18}_{-0.18}\times10^{-6}$ & $(1, \pm 0.50, \mp 1)$ & $1.01^{+0.24}_{-0.21}\times10^{-5}$ & $2.11^{+0.60}_{-0.50}$ \\
 		\hline \hline
 		
 		\multirow{2}{*}{Model} &
 		\multirow{2}{*}{\shortstack{Structure}} & \multicolumn{4}{c}{$B^0\to K^*\nu\bar{\nu}$ Optimization}\\
 		\cline{3-6}
 		& & $\mathcal{B}_{\text{max}}$ & Param.\ Values $[x, y, z]$ & $\mathcal{B}(B^+\to K^+\nu\bar{\nu})$ & $R$ \\
 		\hline \hline 
 		
 		\multirow{4}{*}{BP-I} 
 		& A & $9.94^{+2.34}_{-2.06}\times10^{-6}$ & $(x, 0, 0)$,$(0, y, z)$ & $4.77^{+0.18}_{-0.18}\times10^{-6}$ & $2.09^{+0.60}_{-0.50}$  \\
 		& B & $1.00^{+0.24}_{-0.21}\times10^{-5}$ & $(1, \pm 1, \mp 0.13)$ & $4.72^{+0.18}_{-0.18}\times10^{-6}$ & $2.12^{+0.60}_{-0.50}$  \\
 		& C & $1.00^{+0.24}_{-0.21}\times10^{-5}$ & $(1, \pm 0.13, \mp 1)$ & $4.72^{+0.18}_{-0.18}\times10^{-6}$ & $2.12^{+0.60}_{-0.50}$  \\
 		& D & $1.02^{+0.24}_{-0.21}\times10^{-5}$ & $(1, \pm 0.54, \mp 1)$ & $5.06^{+0.19}_{-0.19}\times10^{-6}$ & $2.01^{+0.57}_{-0.47}$ \\
 		\hline 
 		
 		\multirow{4}{*}{BP-II} 
 		& A & $1.01^{+0.24}_{-0.21}\times10^{-5}$ & $(1, \pm 1, 0)$ & $4.64^{+0.17}_{-0.17}\times10^{-6}$ & $2.18^{+0.62}_{-0.52}$  \\
 		& B & $9.96^{+2.34}_{-2.06}\times10^{-6}$ & $(1, \pm 1, \pm 0.10)$ & $4.76^{+0.18}_{-0.18}\times10^{-6}$ & $2.09^{+0.59}_{-0.49}$  \\
 		& C & $9.96^{+2.34}_{-2.06}\times10^{-6}$ & $(1, \pm 0.10, \pm 1)$ & $4.76^{+0.18}_{-0.18}\times10^{-6}$ & $2.09^{+0.59}_{-0.49}$  \\
 		& D & $1.02^{+0.24}_{-0.21}\times10^{-5}$ & $(1, \pm 0.56, \mp 1)$ & $4.80^{+0.18}_{-0.18}\times10^{-6}$ & $2.12^{+0.60}_{-0.50}$ \\
 		\hline \hline
 	\end{tabular}
 \end{table}
 
With the allowed parameter space established, we now determine the upper bounds on the branching ratios $\mathcal{B}(B^+\to K^+\nu\bar{\nu})$ and $\mathcal{B}(B^0\to K^{*0}\nu\bar{\nu})$, as well as the predicted ratio $R$. Independent numerical optimizations are carried out for each benchmark model across structures A--D, seeking to maximize either the charged or the neutral decay channel. The resulting optimal values, along with the form-factor uncertainties, are compiled in Table~\ref{tab:max_charged}. For each case, we report the global maximum $\mathcal{B}_{\text{max}}$, the parameter configuration $[x, y, z]$ at which it occurs, the predicted branching fraction for the complementary channel, and the associated value of $R$.

In the case of BP-I-A, we find that NP contributions tend to suppress the neutral mode $B^0 \to K^{*0}\nu\bar{\nu}$. Consequently, its maximum branching ratio corresponds to the SM limit\footnote{Minor numerical differences between our calculated SM rates in Table~\ref{tab:max_charged} and the predictions in Ref.~\cite{Becirevic:2023aov} arise from slight variations in parametric inputs (such as CKM elements and lifetimes) and the form-factor parametrization.}, represented by parameter configurations such as $(x, 0, 0)$, where the vanishing quark couplings ($y=z=0$) render the lepton coupling $x$ irrelevant. The opposite scenario is $(0, y, z)$, where the lepton coupling vanishes and all effects come solely from the quark couplings.
 Conversely, maximizing the $B^+\to K^+\nu\bar{\nu}$ rate yields a modest $0.21\%$ enhancement in the central value over the SM prediction; at this point, the predicted rate for the neutral mode drops slightly below its SM value. This anti-correlation indicates that a simultaneous maximization of both decay modes is not possible under flavor structure A. A similar anti-correlation is observed for BP-I-B and BP-I-C, though a small $0.60\%$ enhancement in the central value can be achieved for the neutral mode. 

Flavor structure D (BP-I-D), however, presents a distinct phenomenology: NP contributions can provide constructive interference for both channels, allowing for maximal central-value enhancements of $9.85\%$ and $2.62\%$ for the charged and neutral modes, respectively. Notably, at the parameter point where the neutral mode is maximized, the charged mode is also enhanced by $6.08\%$ relative to the SM, leading to a $3.83\%$ reduction in the central value of $R$. However, as one continues maximizing the charged mode up to $5.24\times10^{-6}$, the predicted rate for the neutral mode drops to $9.71\times10^{-6}$, indicating again an anti-correlated behavior between the two decay channels.

Turning to the fermion-rich model (BP-II), structure A exhibits the inverse behavior of the scalar-rich model: NP contributions tend to suppress the charged mode $B^+\to K^+\nu\bar{\nu}$, limiting its maximum to the SM value of $4.77\times 10^{-6}$ via configurations such as $(x, 0, 0)$ or $(0, y, z)$. When maximizing the neutral mode $B^0 \to K^{*0}\nu\bar{\nu}$ to $1.01\times 10^{-5}$ (a $1.61\%$ enhancement), the charged mode drops to $4.64\times 10^{-6}$, which increases the ratio $R$ to $2.18$. Structures B and C in BP-II follow the same anti-correlation observed in BP-I, where any enhancement in one mode forces the complementary mode below its SM expectation.

In contrast, structure D (BP-II-D) allows for constructive interference in both channels simultaneously. Specifically, at the parameter point that maximizes the neutral mode ($\mathcal{B}_{\text{max}} = 1.02\times 10^{-5}$, a $2.62\%$ enhancement), the charged mode concurrently reaches $4.80\times 10^{-6}$ (a $0.63\%$ enhancement), which matches the maximum value obtained in the dedicated charged-mode scan. In the reverse optimization, the neutral-mode rate at the charged maximum reaches $1.01\times 10^{-5}$. Because the relative enhancement in the neutral channel exceeds that in the charged channel, the central value of $R$ increases from $2.09$ to $2.12$.

As listed in Table~\ref{tab:max_charged}, the absolute branching fractions across all flavor structures (A--D) and both benchmark models are subject to hadronic form-factor uncertainties. These uncertainties are nearly uniform across all entries, amounting to roughly $\pm 4\%$ for $\mathcal{B}(B^+\to K^+\nu\bar{\nu})$, $\pm 20\%$ for $\mathcal{B}(B^0\to K^{*0}\nu\bar{\nu})$, and $\pm 25\%$ for the ratio $R$, largely dominated by the hadronic matrix elements in the $B\to K^*$ transition. Taking these theoretical uncertainties into account, the lowest attainable ratio across all investigated scenarios can be $R \approx 1.41$, realized in $\text{BP-I-D}$ under the $B^+\to K^+\nu\bar{\nu}$ optimization.

We conclude this section by considering the simplest flavor structure, namely $g^3_1 = g^2_1 = g^3_{1'} = g^2_{1'}$. As is evident from Eqs.~\eqref{eq:SRCld} and \eqref{eq:SRClq}, the scalar-rich model under the BP-I mass scheme yields $[C_{\ell q}^{(1)}]_{\ell\ell23} = -[C_{\ell d}]_{\ell\ell23}$, or equivalently, $\Delta C_L^{\nu_\ell} = -\Delta C_R^{\nu_\ell}$. Because the $B^+\to K^+\nu\bar{\nu}$ amplitude depends on the combination $(\Delta C_L^{\nu_\ell} + \Delta C_R^{\nu_\ell})$, this exact cancellation implies that NP contributions to the charged mode strictly vanish, leaving its branching fraction at the SM prediction. Conversely, for the fermion-rich model under the BP-II mass scheme, this same flavor structure results in $\Delta C_L^{\nu_\ell} = \Delta C_R^{\nu_\ell}$. Nevertheless, because the relevant loop function $J_4$ is negative, the resulting NP loop effects contribute destructively, suppressing both decay modes below their SM expectations.

\section{Conclusion}
\label{sec:Summary}

The recent evidence for $B^+ \to K^+ \nu \bar{\nu}$ reported by Belle II, combined with the existing constraints on $B^0 \to K^{*0} \nu \bar{\nu}$, implies a deviation from the SM prediction that necessitates NP contributions generating both left-handed ($\mathcal{O}_V^L$) and right-handed ($\mathcal{O}_V^R$) effective operators. 
Adopting a systematic bottom-up approach, we classified the renormalizable one-loop topologies capable of generating these operators, and identified the irreducible box diagram as the unique candidate for generating the relevant operators at the one-loop level without inducing tree-level contributions. Guided by gauge symmetry conservation, we cataloged the allowed quantum numbers for the internal mediators, and proposed two minimal benchmark models for detailed study: a scalar-rich model, constructed by combining diagrams that independently generate $\mathcal{O}_V^R$ and $\mathcal{O}_V^L$, and a fermion-rich model, where a single diagram topology is versatile enough to generate both operator chiralities.

We subjected these models to a comprehensive phenomenological analysis, constraining the parameter space using a complementary set of flavor observables and precision electroweak data. Under the assumption that NP couples exclusively to the third lepton generation, our numerical evaluation reveals that the parameter space for both benchmark models is severely constrained, primarily by $B_s - \bar{B}_s$ mixing and $b \to s \gamma$. 
Within this surviving parameter space, we find that the specific flavor structure directly dictates the pattern of loop-induced deviations in the two decay modes. For the correlated structures (A, B, and C), the loop contributions to the charged and neutral channels are strongly anticorrelated. Structure D, however, provides enough flexibility to generate constructive loop interference, simultaneously enhancing both decay modes. 
Due to the asymmetric enhancements, the ratio $R$ decreases in the scalar-rich model, where the charged-mode enhancement dominates, and increases in the fermion-rich model, where the neutral mode is more strongly enhanced.

\section*{Acknowledgments}
This work is supported by the National Natural Science Foundation of China under Grants No. 12135006 and No. 12275067, 
the Natural Science Foundation of Henan Province under Grant No. 242300421390, 
the Science and Technology R\&D Program Joint Fund Project of Henan Province under Grant No. 225200810030, 
the Science and Technology Innovation Leading Talent Support Program of Henan Province under Grant No. 254000510039, 
as well as the National Key R\&D Program of China under Grant No. 2023YFA1606000. 

\appendix
\counterwithin*{equation}{section}
\renewcommand\theequation{\thesection\arabic{equation}}

\section{Loop-Integral Functions}
\label{app:loopfunc}

This appendix provides the explicit analytical expressions for the loop-integral functions referenced in the main text.

The evaluation of the NP contributions in the benchmark models (Sec.~\ref{sec:benchmark-model}) utilizes the following four-point integrals:
\begin{align}
	I_{4} (m_1, m_2, m_3, m_4) 
	&\equiv \int \frac{\mathrm{d}^{d} k}{(2\pi)^{d} i} \frac{1}{(k^{2} - m_1^{2})(k^{2} - m_2^{2})(k^{2} - m_3^{2})(k^{2} - m_4^{2})} \nonumber \\
	&= \frac{1}{(4\pi)^{2}} \left[ \frac{m_1^{2}}{(m_1^{2} - m_2^{2})(m_1^{2} - m_3^{2})(m_1^{2} - m_4^{2})} \ln\frac{m_4^{2}}{m_1^{2}} \right. \nonumber \\
	&\qquad + \frac{m_2^{2}}{(m_2^{2} - m_1^{2})(m_2^{2} - m_3^{2})(m_2^{2} - m_4^{2})} \ln\frac{m_4^{2}}{m_2^{2}} \nonumber \\
	&\qquad + \left. \frac{m_3^{2}}{(m_3^{2} - m_1^{2})(m_3^{2} - m_2^{2})(m_3^{2} - m_4^{2})} \ln\frac{m_4^{2}}{m_3^{2}} \right], \\	
	J_{4} (m_1, m_2, m_3, m_4) 
	&\equiv \int \frac{\mathrm{d}^{d} k}{(2\pi)^{d} i} \frac{k^{2}}{(k^{2} - m_1^{2})(k^{2} - m_2^{2})(k^{2} - m_3^{2})(k^{2} - m_4^{2})} \nonumber \\
	&= \frac{1}{(4\pi)^{2}} \left[ \frac{m_1^{4}}{(m_1^{2} - m_2^{2})(m_1^{2} - m_3^{2})(m_1^{2} - m_4^{2})} \ln\frac{m_4^{2}}{m_1^{2}} \right. \nonumber \\
	&\quad + \frac{m_2^{4}}{(m_2^{2} - m_1^{2})(m_2^{2} - m_3^{2})(m_2^{2} - m_4^{2})} \ln\frac{m_4^{2}}{m_2^{2}} \nonumber \\
	&\quad + \left. \frac{m_3^{4}}{(m_3^{2} - m_1^{2})(m_3^{2} - m_2^{2})(m_3^{2} - m_4^{2})} \ln\frac{m_4^{2}}{m_3^{2}} \right].
\end{align}

Additionally, the analysis of the radiative decay $b \to s\gamma$ in Sec.~\ref{sec:bsgamma} involves the auxiliary loop functions $F_7(x)$ and $\widetilde{F}_7(x)$, which are defined as:
\begin{align}
	F_7(x) & = \frac{x^3 - 6x^2 + 3x + 6x\log x + 2}{12(x-1)^4}, \nonumber\\
	\widetilde{F}_7(x) & = x^{-1}F_7(x^{-1}). 
\end{align}

\section{One-loop Corrections to $Z \bar{f}_i f_j$  Couplings}
\label{app:Zcoupling}

In this section, we revisit the radiative corrections to the couplings between the $Z$ boson and two SM fermions, $f_i$ and $f_j$, induced by the new physics sector. We consider general momentum transfer $q^2$ to accommodate both on-shell $Z$ decays and off-shell exchange processes. The effective Lagrangian governing the $Z \bar{f}_i f_j$ interaction is defined as
\begin{align}
	\mathcal{L}_{\text{eff}} = \frac{g_2}{c_W}\bar{f}_i\gamma^\mu\left[g_L^{ij}(q^2)P_L + g_R^{ij}(q^2)P_R\right]f_j Z_\mu + \mathrm{h.c.}, 
\end{align}
where $q$ is the four-momentum of the $Z$ boson, $g_2$ is the $SU(2)_L$ gauge coupling, and $c_W \equiv \cos \theta_W$, with $\theta_W$ being the weak mixing angle. The form factors are decomposed into the tree-level SM contribution and the new physics loop correction:
\begin{align}
	g_{L(R)}^{ij}(q^2) &= g_{L(R)}^{\text{SM}}\delta_{ij} + \Delta g_{L(R)}^{ij}(q^2)\,.
\end{align}
The tree-level SM couplings are given by
\begin{align}
	g_L^{\text{SM}} &= T^3_{f_i} - Q_{f_i} s_W^2\,, \quad g_R^{\text{SM}} = -Q_{f_i} s_W^2\,, 
\end{align}
where $T^3_{f}$ is the third component of weak isospin and $Q_{f}$ is the electric charge.

The relevant Feynman diagrams contributing to these corrections are depicted in Fig.~\ref{fig:Zff}. The interactions of the $Z$ boson with the new scalars and fermions below the electroweak scale are parameterized by the following Lagrangian:
\begin{align}
	\label{eq:Lz}
	\mathcal{L}_Z &= \frac{g_2}{c_W}Z_\mu\left(\bar{\Psi}_A\gamma^\mu\left[g_{A,B}^{\Psi,L}P_L + g_{A,B}^{\Psi,R}P_R\right]\Psi_B + g_{MN}^{\Phi} \Phi_M^{\dagger} i\overset{\leftrightarrow}{\partial^\mu}\Phi_N\right) + \mathrm{h.c.}\,, 
\end{align}  
where we utilize the notation $a\overset{\leftrightarrow}{\partial^{\mu}}b \equiv a(\partial^{\mu}b)-(\partial^{\mu}a)b$. The indices $A, B$ (for fermions $\Psi$) and $M, N$ (for scalars $\Phi$) implicitly run over all relevant internal degrees of freedom, including flavor, color, and $SU(2)$ indices. The Yukawa interactions linking a new scalar $\Phi_M$ and a new fermion $\Psi_A$ to a SM fermion $f_i$ are given by
\begin{align}
	\label{eq:Yukawa}
	\mathcal{L}_{\mathrm{int}} =\bar{\Psi}_A \left( L_{AM}^{f_i} P_L +R_{AM}^{f_i} P_R \right)f_i  \Phi_M  + \mathrm{h.c.}
\end{align}

We calculate the loop contributions to the form factors $\Delta g_{L,R}^{ij}(q^2)$ using the Lagrangians in Eqs.~\eqref{eq:Lz} and \eqref{eq:Yukawa}. The calculation is performed with the aid of \textsc{Package-X} 2.1.1~\cite{Patel:2015tea, Patel:2016fam}, adopting the covariant derivative convention and Feynman rules from Ref.~\cite{Crivellin:2021ejk}.
In our evaluation, we adopt the limit of vanishing external SM fermion masses ($m_{i,j} \to 0$) for both vertex and external leg corrections. An exception is made for the contributions proportional to $\delta_{AB}\delta_{MN}$ involving the denominator $(m_j^2 - m_i^2)$, where setting the masses directly to zero would result in a kinematic singularity. We therefore retain the mass dependence in these specific terms to regulate the expression. We observe that in the limit where the tree-level couplings are flavor-diagonal ($g_{f_{iL}}^{\text{SM}}=g_{f_{jL}}^{\text{SM}}$)—as holds true in the SM—this singularity cancels out exactly, yielding a finite result consistent with the massless approximation.

The resulting left-handed contribution $\Delta g_{L}^{ij}(q^2)$ is found to be:
\begin{align}
	\Delta g_{L}^{ij}(q^2) =& -\frac{\chi_Z L^{i*}_{AM}L^{j}_{BN}}{16\pi^2}\Bigg\{ \bigg[
	g_{f_{jL}}^{\text{SM}}\frac{m_i^2}{m^2_j-m_i^2}B_1(m_i^2;M_{\Psi_A},M_{\Phi_N}) \nonumber \\
	&-g_{f_{iL}}^{\text{SM}}\frac{m_j^2}{m^2_j-m_i^2}B_1(m_j^2;M_{\Psi_A},M_{\Phi_N}) 	 
	\bigg]\delta_{AB}\delta_{MN} \nonumber \\
	&+\bigg[g^L_{AB}M_{\Psi_A}M_{\Psi_B}C_0(0,q^2,0; M_{\Phi_N},M_{\Psi_B},M_{\Psi_A}) \nonumber \\
	&- g^R_{AB}\bigg(B_0(q^2;M_{\Psi_B},M_{\Psi_A}) +M^2_{\Phi_N}C_0(0,q^2,0;M_{\Phi_N},M_{\Psi_B},M_{\Psi_A}) \nonumber \\
	&-2 C_{00}(0,q^2,0;M_{\Phi_N},M_{\Psi_B},M_{\Psi_A})\bigg) \bigg]\delta_{MN} \nonumber \\
	&+2g^{\Phi}_{MN}C_{00}(0,q^2,0;,M_{\Psi_A},M_{\Phi_N},M_{\Phi_M})\delta_{AB}\Bigg\}\,,
\end{align}
and the right-handed contribution follows via the exchange of chirality parameters:
\begin{align}
	\Delta g_{R}^{ij}(q^2) = \Delta g_{L}^{ij}(q^2) \big(L \leftrightarrow R\big)\,.
\end{align}
Here, $B_{0,1}$, $C_0$, and $C_{00}$ denote the Passarino-Veltman scalar integrals~\cite{Patel:2015tea}, and $\chi_Z$ represents the relevant group theory factor (e.g., color factor).

For phenomenological applications where the new physics scale is significantly heavier than the momentum transfer, it is useful to expand the results in powers of $q^2$. Up to the first order in $q^2$, the form factors are
\begin{align}\label{eq:DeltagL}
	\Delta g_{L}^{ij}(q^2) = & \frac{\chi_Z L^{i*}_{AM}L^{j}_{BN}}{32\pi^2}\Bigg\{
	2g^L_{AB}\delta_{MN}\frac{M_{\Psi_B}M_{\Psi_A}}{M^2_{\Phi_M}}G_Z(x_{AM},x_{BM}) \nonumber \\[0.02cm]
	&-g^R_{AB}\delta_{MN}F_Z(x_{AM},x_{BM},M_{\Phi_M}) 
	-g^{\Phi}_{MN}\delta_{AB} H_Z(x_{AM},x_{BM},M_{\Psi_A}) \nonumber \\[0.02cm]
	&-g_{f_{iL}}^{\text{SM}}\delta_{AB}\delta_{MN}I_Z(x_{AM},M_{\Phi_M}) \nonumber \\[0.02cm]
	&+q^2\bigg[   g^L_{AB}\delta_{MN}\frac{M_{\Psi_B}M_{\Psi_A}}{M^4_{\Phi_M}}\tilde{G}_Z(x_{AM},x_{BM})
	-\frac{2}{3}\frac{g^R_{AB}\delta_{MN}}{M^2_{\Phi_M}}\tilde{F}_Z(x_{AM},x_{BM})\nonumber \\[0.02cm]
	&+\frac{g^{\Phi}_{MN}\delta_{AB} }{3M^2_{\Psi_A}}\tilde{H}_Z(x_{AM},x_{BM})
	\bigg]
	\Bigg\}\,,
\end{align}
where we have defined the mass ratios $x_{AM} \equiv M_{\Psi_A}^2/M_{\Phi_M}^2$. The loop functions are defined as:
\begin{align}
	G_Z(x, y) &= xF_V(x, y) + (x \leftrightarrow y), \nonumber \\
	F_Z(x, y, m) & = \left[x^2 F_V(x, y) + (x \leftrightarrow y)\right] -\frac{1}{2}- \overline{\mathrm{div}}_\varepsilon, \nonumber \\
	H_Z(x, y, m) & = \left[yF_V(x, y) + (x \leftrightarrow y)\right] + \frac{3}{2} + \overline{\mathrm{div}}_\varepsilon, \nonumber \\
	I_Z(x, m) & = \frac{3x-1}{2(x-1)} - x^2 F_V(x, 1) + \overline{\mathrm{div}}_\varepsilon, \nonumber \\
	\widetilde{G}_Z(x, y) &= xK_V(x, y) + (x \leftrightarrow y), \nonumber \\
	\widetilde{F}_Z(x, y) &= \left[ x^2 K_V(x, y) - \frac{x^2}{x-y}F_V(x, y) \right] + (x \leftrightarrow y), \nonumber \\
	\widetilde{H}_Z(x, y) &= \left[ \frac{x^2 y}{(y-1)(x-y)^2} - \frac{x^2 y^2(3x - y - 2) \log(x)}{(x-1)^2(x-y)^3} \right] + (x \leftrightarrow y), 
\end{align}
where the divergence parameter is given by $\overline{\mathrm{div}}_\varepsilon = \Delta_\varepsilon - \log \left( m^2/\mu^2 \right)$ in dimensional regularization. The auxiliary functions $F_V$ and $K_V$ are
\begin{equation}
	F_V(x, y) = \frac{\log(x)}{(x-1)(x-y)}, \qquad K_V(x, y) = \frac{(x^2 + xy - 2y) \log(x)}{(x-1)^2(x-y)^3} - \frac{1}{(x-1)(x-y)^2}\,.
\end{equation}

We note that our analytical results differ slightly from those reported in Ref.~\cite{Arnan:2019uhr}. Specifically, we observe an overall sign difference in the terms proportional to $g^{\Phi}_{MN}$, along with minor discrepancies in the finite parts of the loop functions $F_Z$, $H_Z$, and $I_Z$. However, the physical behavior at $q^2=0$ remains consistent. Since the NP sector introduces no new sources of electroweak symmetry breaking, the contributions from self-energies exactly cancel those from vertex corrections at zero momentum transfer. We confirm that our expressions for $\Delta g_{L,R}^{ij}(q^2)$ vanish at $q^2=0$, in agreement with the limiting behavior established in Ref.~\cite{Arnan:2019uhr}.

\section{Quantum numbers of the internal fields and UV completions}
\label{app:models}
 
In Sec.~\ref{sec:classification}, we derived physical model diagrams from the underlying topologies by imposing SM gauge invariance. While the allowed quantum numbers for diagram (c) (Fig.~\ref{fig:box}) were presented in Table~\ref{tab:T1-c-L} as a representative example, this appendix provides the complete classification for the remaining five topologies (Tables~\ref{tab:T1-a}--\ref{tab:T1-f}).

Connecting the high-scale UV models to the LEFT basis necessitates an intermediate matching onto the dimension-six SMEFT operators. Consequently, we calculate the Wilson coefficients for the relevant SMEFT operators generated by each diagram across all valid SU(2)$_L$ assignments. The relative strengths of these contributions are summarized in Tables~\ref{tab:T1-a-R-Mod}--\ref{tab:T1-f-L-Mod}.

Our calculation involves two distinct SU(2)$_L$ contraction structures: vertices involving two doublets and an anti-triplet, and vertices involving a doublet, a triplet, and an anti-doublet (along with their Hermitian conjugates). To facilitate the evaluation of these contractions, we utilize two SU(2)$_L$ tensors, $\xi^A$ and $\chi^A$. The tensor $\xi^A$, defined in the main text, mediates interactions between two doublets and a triplet. For the second case, we introduce the tensor $\chi^A$, defined as:
\begin{align}
	\chi^A = \left( (\tau^1+i \tau^2)/2, \, -\tau^3/\sqrt{2}, \, (-\tau^1+i \tau^2)/2 \right), \qquad A\in \{1,2,3\},
\end{align}
where $\tau^i$ are the Pauli matrices:
\begin{align*}
	\tau^1 = \begin{pmatrix} 0 & 1 \\ 1 & 0 \end{pmatrix},\quad
	\tau^2 = \begin{pmatrix} 0 & -i \\ i & 0 \end{pmatrix},\quad
	\tau^3 = \begin{pmatrix} 1 & 0 \\ 0 & -1 \end{pmatrix}.
\end{align*}
The tensor $\chi^A$ and its conjugate $\chi^{A\dagger}$ govern the interaction vertices involving one doublet, one triplet, and one anti-doublet. In all cases, the index $A$ corresponds to the SU(2)$_L$ index of the triplet or anti-triplet field.

To illustrate the application of the $\chi^A$ tensor, we consider Model A-VII based on diagram (c) (see Table~\ref{tab:T1-c-L}). Since the scalars $X_2$ and $X_4$ share identical SM quantum numbers in this model, we identify them as a single physical field ($X_2 = X_4$). The explicit interaction Lagrangian, including the full SU(2)$_L$ structure, is then given by:
\begin{align}
	\mathcal{L} &\supset g^t_{1}\bar{\Psi}_{1}Q_{t} X_{2}+g^{p}_{2}\bar{L}_{p}\Psi_{3} X^{\dagger}_{2}
	+g^{r*}_{2}\bar{\Psi}_{3} L_{r}X_{2}+g^{s*}_{1}\bar{Q}_s \Psi_{1} X^{\dagger}_{2} \nonumber \\
	&= g^t_{1}\bar{\Psi}_{1\alpha}Q_{t\alpha'} X^{A}_{2}\chi^A_{\alpha\alpha'}
	+g^{p}_{2}\bar{L}_{p\beta}\Psi_{3\beta'} X^{*B}_{2}(\chi^{B\dagger})_{\beta\beta'} \nonumber \\
	&\quad +g^{r*}_{2}\bar{\Psi}_{3\rho} L_{r\rho'}X^{A'}_{2}\chi^{A'}_{\rho\rho'}
	+g^{s*}_{1}\bar{Q}_{s\sigma} \Psi_{1\sigma'} X^{*B'}_{2}(\chi^{B'\dagger})_{\sigma\sigma'}\,,
\end{align}
where the Greek indices $\alpha, \beta, \rho, \sigma$ run over the SU(2)$_L$ doublet components $1, 2$.

\begin{table}[!tbph] 
	\renewcommand\arraystretch{1.3}  
	\centering 
	\tabcolsep=0.38cm
	\caption{Possible assignments of quantum numbers under the $\text{SU(3)}_{c}\otimes \text{SU}(2)_L\otimes \text{U}(1)_{\alpha}$ symmetry to the mediators in diagram ($a$) to generate the operator $\mathcal{O}^R_V$. }
	\begin{tabular}{c|cccc}
		\hline \hline
		SU(3)$_c$ & $\Psi_1$    & $X_2$  & $\Psi_3$ & $X_4$  \\
		\hline
		A & 1 & $\bar{3}$ & 1 & 1 \\
		B & 3 & 1 & 3 & 3 \\
		\hline \hline
		SU(2)$_L$ & $\Psi_1$    & $X_2$  & $\Psi_3$ & $X_4$  \\
		\hline
		I & 1 & 1 & 1 & 2 \\
		II & 2 & 2 & 2 & 1 \\
		III & 2 &2 & 2 & 3 \\
		IV & 3 & 3 & 3 & 2 \\
		\hline \hline
		U(1)$_Y$ & $\Psi_1$    & $X_2$  & $\Psi_3$ & $X_4$  \\
		\hline
		i & $\alpha$ & $\alpha+1/3$ & $\alpha$ & $\alpha-1/2$ \\
		\hline \hline
	\end{tabular}
	\label{tab:T1-a}   
\end{table}

\begin{table}[!tbph] 
	\renewcommand\arraystretch{1.3}  
	\centering 
	\tabcolsep=0.38cm
	\caption{Possible assignments of quantum numbers under the $\text{SU(3)}_{c}\otimes \text{SU}(2)_L\otimes \text{U}(1)_{\alpha}$ symmetry to the mediators in diagram ($b$) to generate operator $\mathcal{O}^R_V$. }
	\begin{tabular}{c|cccc}
		\hline \hline
		SU(3)$_c$ & $\Psi_1$    & $X_2$  & $\Psi_3$ & $X_4$  \\
		\hline
		A & 1 & $\bar{3}$ & $\bar{3}$ & 1 \\
		B & 3 & 1 & 1 & 3 \\
		\hline \hline
		SU(2)$_L$ & $\Psi_1$    & $X_2$  & $\Psi_3$ & $X_4$  \\
		\hline
		I & 1 & 1 & 2 & 2 \\
		II & 2 & 2 & 1 & 1 \\
		III & 2 & 2 & 3 & 3 \\
		IV & 3 &3 & 2 & 2 \\
		\hline \hline
		U(1)$_Y$ & $\Psi_1$    & $X_2$  & $\Psi_3$ & $X_4$  \\
		\hline
		i & $\alpha$ & $\alpha+1/3$ & $\alpha-1/6$ & $\alpha-1/2$ \\
		\hline \hline
	\end{tabular}
	\label{tab:T1-b}   
\end{table}

\begin{table}[!h] 
	\renewcommand\arraystretch{1.3}  
	\centering 
	\tabcolsep=0.38cm
	\caption{Possible assignments of quantum numbers under the $\text{SU(3)}_{c}\otimes \text{SU}(2)_L\otimes \text{U}(1)_{\alpha}$ symmetry to the mediators in diagram ($c$) to generate operator $\mathcal{O}^R_V$. }
	\begin{tabular}{c|cccc}
			\hline \hline
			SU(3)$_c$ & $\Psi_1$    & $X_2$  & $\Psi_3$ & $X_4$  \\
			\hline
			A & 1 & $\bar{3}$ & $\bar{3}$ & $\bar{3}$ \\
			B & 3 & 1 & 1 & 1 \\
			\hline \hline
			SU(2)$_L$ & $\Psi_1$    & $X_2$  & $\Psi_3$ & $X_4$  \\
			\hline
			I & 1 & 1 & 2 & 1 \\
			II & 2 & 2 & 1 & 2 \\
			III & 2 &2 & 3 & 2 \\
			IV & 3 & 3 & 2 &3 \\
			\hline \hline
			U(1)$_Y$ & $\Psi_1$    & $X_2$  & $\Psi_3$ & $X_4$  \\
			\hline
			i & $\alpha$ & $\alpha+1/3$ & $\alpha-1/6$ & $\alpha+1/3$ \\
			\hline \hline
		\end{tabular}
	\label{tab:T1-c-R}   
\end{table}

\begin{table}[h] 
	\renewcommand\arraystretch{1.3}  
	\centering 
	\tabcolsep=0.38cm
	\caption{Possible assignments of quantum numbers under the $\text{SU(3)}_{c}\otimes \text{SU}(2)_L\otimes \text{U}(1)_{\alpha}$ symmetry to the mediators in diagram ($d$) to generate operator $\mathcal{O}^L_V$. }
		\begin{tabular}{c|cccc}
		\hline \hline
		SU(3)$_c$ & $\Psi_1$    & $X_2$  & $\Psi_3$ & $X_4$  \\
		\hline
		A & 1 & $\bar{3}$ & $\bar{3}$ & $\bar{3}$ \\
		B & 3 & 1 & 1 & 1 \\
		\hline \hline
		SU(2)$_L$ & $\Psi_1$    & $X_2$  & $\Psi_3$ & $X_4$  \\
		\hline
		I & 1 & 2 & 1 & 2 \\
		II & 1 & 2 & 3 & 2 \\
		III & 2 &1 & 2 & 1 \\
		IV & 2 & 1 & 2 &3 \\
		V & 3 & 2& 1 & 2\\
		VI & 3 & 2 & 3 & 2\\
		VII & 2 &3 & 2 & 1 \\
		VIII & 2 & 3 & 2 &3 \\
		\hline \hline
		U(1)$_Y$ & $\Psi_1$    & $X_2$  & $\Psi_3$ & $X_4$  \\
		\hline
		i & $\alpha$ & $\alpha-1/6$ & $\alpha+1/3$ & $\alpha-1/6$ \\
		\hline \hline
	\end{tabular}
	\label{tab:T1-d-L}   
\end{table}

\begin{table}[h] 
	\renewcommand\arraystretch{1.3}  
	\centering 
	\tabcolsep=0.38cm
	\caption{Possible assignments of quantum numbers under the $\text{SU(3)}_{c}\otimes \text{SU}(2)_L\otimes \text{U}(1)_{\alpha}$ symmetry to the mediators in diagram ($d$) to generate operator $\mathcal{O}^R_V$. }
		\begin{tabular}{c|cccc}
		\hline \hline
		SU(3)$_c$ & $\Psi_1$    & $X_2$  & $\Psi_3$ & $X_4$  \\
		\hline
		A & 1 & $\bar{3}$ & $\bar{3}$ & $\bar{3}$ \\
		B & 3 & 1 & 1 & 1 \\
		\hline \hline
		SU(2)$_L$ & $\Psi_1$    & $X_2$  & $\Psi_3$ & $X_4$  \\
		\hline
		I & 1 & 1 & 2 & 1 \\
		II & 2 & 2 & 1 & 2 \\
		III & 2 &2 & 3 & 2 \\
		IV & 3 & 3 & 2 &3 \\
		\hline \hline
		U(1)$_Y$ & $\Psi_1$    & $X_2$  & $\Psi_3$ & $X_4$  \\
		\hline
		i & $\alpha$ &$\alpha+1/3$ & $\alpha+5/6$ & $\alpha+1/3$ \\
		\hline \hline
	\end{tabular}
	\label{tab:T1-d-R}   
\end{table}

\begin{table}[h] 
	\renewcommand\arraystretch{1.3}  
	\centering 
	\tabcolsep=0.38cm
	\caption{Possible assignments of quantum numbers under the $\text{SU(3)}_{c}\otimes \text{SU}(2)_L\otimes \text{U}(1)_{\alpha}$ symmetry to the mediators in diagram ($e$) to generate operator $\mathcal{O}^L_V$. }
	\begin{tabular}{c|cccc}
		\hline \hline
		SU(3)$_c$ & $\Psi_1$    & $X_2$  & $\Psi_3$ & $X_4$  \\
		\hline
		A & 1 & $\bar{3}$ & 1 & 1 \\
		B & 3 & 1 & 3 &3 \\
		\hline \hline
		SU(2)$_L$ & $\Psi_1$    & $X_2$  & $\Psi_3$ & $X_4$  \\
		\hline
		I & 1 & 2 & 1 & 2 \\
		II & 1 & 2 & 3 & 2 \\
		III & 2 & 1 & 2 &3 \\
		IV & 3 & 2& 1 & 2\\
		V & 3 & 2 & 3 & 2\\
		VI & 2 &3 & 2 & 1 \\
		VII & 2 & 3 & 2 &3 \\
		\hline \hline
		U(1)$_Y$ & $\Psi_1$    & $X_2$  & $\Psi_3$ & $X_4$  \\
		\hline
		i & $\alpha$ & $\alpha-1/6$ & $\alpha$ & $\alpha+1/2$ \\
		\hline \hline
	\end{tabular}
	\label{tab:T1-e}   
\end{table}

\begin{table}[h] 
	\renewcommand\arraystretch{1.3}  
	\centering 
	\tabcolsep=0.38cm
	\caption{Possible assignments of quantum numbers under the $\text{SU(3)}_{c}\otimes \text{SU}(2)_L\otimes \text{U}(1)_{\alpha}$ symmetry to the mediators in diagram ($f$) to generate operator $\mathcal{O}^L_V$. }
	\begin{tabular}{c|cccc}
		\hline \hline
		SU(3)$_c$ & $\Psi_1$    & $X_2$  & $\Psi_3$ & $X_4$  \\
		\hline
		A & 1 & $\bar{3}$ & $\bar{3}$ & 1 \\
		B & 3 & 1 & 1 & 3 \\
		\hline \hline
		SU(2)$_L$ & $\Psi_1$    & $X_2$  & $\Psi_3$ & $X_4$  \\
		\hline
		I & 1 & 2 & 1 & 2 \\
		II & 1 & 2 & 3 & 2 \\
		III & 2 & 1 & 2 &3 \\
		IV & 3 & 2& 1 & 2\\
		V & 3 & 2 & 3 & 2\\
		VI & 2 &3 & 2 & 1 \\
		VII & 2 & 3 & 2 &3 \\
		\hline \hline
		U(1)$_Y$ & $\Psi_1$    & $X_2$  & $\Psi_3$ & $X_4$  \\
		\hline
		i & $\alpha$ & $\alpha-1/6$ & $\alpha+1/3$ & $\alpha+1/2$ \\
		\hline \hline
	\end{tabular}
	\label{tab:T1-f}   
\end{table}

\begin{table}[!tbph] 
	\renewcommand\arraystretch{1.6}  
	\centering 
	\tabcolsep=0.6cm % Adjusted spacing for clarity
	\caption{Relative strengths of the SMEFT operator $Q_{\ell d}$ generated by diagram ($a$) for the different SU(2)$_L$ models. An overall normalization factor, which includes the loop function and couplings, is not presented.}
	\begin{tabular}{c|cccc} 
		\hline\hline
		\diagbox[width=7.5em]{Operator}{ Model} & I & II & III & IV \\
		\hline
		$Q_{\ell d}$ & -1 & 1 & -3/2 & -3/2  \\
		\hline\hline
	\end{tabular}
	\label{tab:T1-a-R-Mod}   
\end{table}

\begin{table}[!tbph] 
	\renewcommand\arraystretch{1.6}  
	\centering 
	\tabcolsep=0.6cm % Adjusted spacing for clarity
	\caption{Relative strengths of the SMEFT operator $Q_{\ell d}$ generated by diagram ($b$) for the different SU(2)$_L$ models. An overall normalization factor, which includes the loop function and couplings, is not presented.}
	\begin{tabular}{c|cccc} 
		\hline\hline
		\diagbox[width=7.5em]{Operator}{ Model} & I & II & III & IV \\
		\hline
		$Q_{\ell d}$ & -1 & -1 & -3/2 & -3/2  \\
		\hline\hline
	\end{tabular}
	\label{tab:T1-b-R-Mod}   
\end{table}

\begin{table}[!tbph] 
	\renewcommand\arraystretch{1.6}  
	\centering 
	\tabcolsep=0.6cm % Adjusted spacing for clarity
	\caption{Relative strengths of the SMEFT operator $Q_{\ell d}$ generated by diagram ($c$) for the different SU(2)$_L$ models. An overall normalization factor, which includes the loop function and couplings, is not presented.}
	\begin{tabular}{c|cccc} 
		\hline\hline
		\diagbox[width=7.5em]{Operator}{ Model} & I & II & III & IV \\
		\hline
		$Q_{\ell d}$ & 1 & 1 & 3/2 & 3/2  \\
		\hline\hline
	\end{tabular}
	\label{tab:T1-c-R-Mod}   
\end{table}

\begin{table}[!tbph] 
	\renewcommand\arraystretch{1.6}  
	\centering 
	\tabcolsep=0.6cm % Adjusted spacing for clarity
	\caption{Relative strengths of the SMEFT operators $Q^{(1)}_{\ell q}$ and $Q^{(3)}_{\ell q}$ generated by diagram ($d$) for the different SU(2)$_L$ models. An overall normalization factor, which includes the loop function and couplings, is not presented.}
	\begin{tabular}{c|cccccccc} 
		\hline\hline
		\diagbox[width=7.5em]{Operator}{ Model} & I & II & III & IV & V & VI & VII & VIII \\
		\hline
		$Q^{(1)}_{\ell q}$ & -1/2 & -3/4 & 1 &   0  & -3/4 & -9/8 &   0  & -3/4 \\
		$Q^{(3)}_{\ell q}$ &  1/2 & -1/4 & 0 &  1/2 &  1/4 &  1/8 & -1/2 &  1/2 \\
		\hline\hline
	\end{tabular}
	\label{tab:T1-d-L-Mod}   
\end{table}

\begin{table}[!tbph] 
	\renewcommand\arraystretch{1.6}  
	\centering 
	\tabcolsep=0.6cm % Adjusted spacing for clarity
	\caption{Relative strengths of the SMEFT operator $Q_{\ell d}$ generated by diagram ($d$) for the different SU(2)$_L$ models. An overall normalization factor, which includes the loop function and couplings, is not presented.}
	\begin{tabular}{c|cccc} 
		\hline\hline
		\diagbox[width=7.5em]{Operator}{ Model} & I & II & III & IV \\
		\hline
		$Q_{\ell d}$ & 1 & -1 & -3/2 & -3/2  \\
		\hline\hline
	\end{tabular}
	\label{tab:T1-d-R-Mod}   
\end{table}

\begin{table}[!tbph] 
	\renewcommand\arraystretch{1.6}  
	\centering 
	\tabcolsep=0.6cm % Adjusted spacing for clarity
	\caption{Relative strengths of the SMEFT operators $Q^{(1)}_{\ell q}$ and $Q^{(3)}_{\ell q}$ generated by diagram ($e$) for the different SU(2)$_L$ models. An overall normalization factor, which includes the loop function and couplings, is not presented.}
	\begin{tabular}{c|ccccccc} 
		\hline\hline
		\diagbox[width=7.5em]{Operator}{ Model} & I & II & III & IV & V & VI & VII \\
		\hline
		$Q^{(1)}_{\ell q}$ & 1 &  0  &  3/4 &  0  & 9/8 &  3/4 & 9/8 \\
		$Q^{(3)}_{\ell q}$ & 0 & 1/2 & -1/4 & 1/2 & 1/8 & -1/4 & 1/8 \\
		\hline\hline
	\end{tabular}
	\label{tab:T1-e-L-Mod}   
\end{table}

\begin{table}[!tbph] 
	\renewcommand\arraystretch{1.6}  
	\centering 
	\tabcolsep=0.6cm % Adjusted spacing for clarity
	\caption{Relative strengths of the SMEFT operators $Q^{(1)}_{\ell q}$ and $Q^{(3)}_{\ell q}$ generated by diagram ($f$) for the different SU(2)$_L$ models. An overall normalization factor, which includes the loop function and couplings, is not presented.}
	\begin{tabular}{c|ccccccc} 
		\hline\hline
		\diagbox[width=7.5em]{Operator}{ Model} & I & II & III & IV & V & VI & VII  \\
		\hline
		$Q^{(1)}_{\ell q}$ &  1/2 & -3/4 &  3/4 & -3/4 & -3/8 & -3/4 &  3/8 \\
		$Q^{(3)}_{\ell q}$ & -1/2 & -1/4 & -1/4 & -1/4 & -5/8 &  1/4 & -5/8 \\
		\hline\hline
	\end{tabular}
	\label{tab:T1-f-L-Mod}   
\end{table}

\bibliographystyle{apsrev4-1}
\bibliography{reference}

\end{document}